\documentclass[bibyear]{aa}
\usepackage[varg]{txfonts}
\usepackage{natbib}
\bibpunct{(}{)}{;}{a}{}{,}

\usepackage{graphicx}
\usepackage{xcolor}
\usepackage{bm}
\usepackage{comment}
\usepackage[colorlinks=true,linkcolor=blue,citecolor=blue,urlcolor=blue]{hyperref}              
\usepackage[capitalise]{cleveref}       
\usepackage{booktabs}   
\usepackage{array}
\usepackage{multirow} 
\usepackage{siunitx}

\newcommand{\der}{\text{d}}

\begin{document}

\title{Imaging without visibilities}
\subtitle{FAST-Effelsberg scintillometry of PSR B1508+55}

\author{
Tim Sprenger\inst{\ref{aff:MPIfR}}\thanks{E-mail: tsprenger@mpifr-bonn.mpg.de} 
\and 
Xun Shi\inst{\ref{aff:SWIFAR}}
\thanks{E-mail: xun@ynu.edu.cn} 
\and 
Olaf Wucknitz\inst{\ref{aff:MPIfR}}
\and 
Robert A. Main\inst{\ref{aff:McGill_Physics},\ref{aff:McGill_TSI}}
}
\institute{
Max-Planck-Institut f\"ur Radioastronomie, Auf dem H\"ugel 69, 53121 Bonn, Germany \label{aff:MPIfR}
\and 
South-Western Institute for Astronomy Research (SWIFAR), Yunnan University, 650500 Kunming, P. R. China
\label{aff:SWIFAR}
\and
Department of Physics, McGill University, 3600 rue University, Montréal, QC H3A 2T8, Canada \label{aff:McGill_Physics}
\and
Trottier Space Institute, McGill University, 3550 rue University, Montréal, QC H3A 2A7, Canada \label{aff:McGill_TSI}
}
\date{\today}

\abstract
{The spatially coherent multipath propagation of pulsar radiation leads to a temporal and spectral interference patterns called scintillation. It is caused by density variations in the ionized interstellar medium, which often take the form of thin scattering screens filled with multiple subimages of the pulsar. PSR B1508+55 is known to be scattered by one or two such screens.}
{We investigate appropriate methods to achieve precise astrometry for a scattering screen from simultaneous observations of only two telescopes on a very long baseline without forming visibilities.}
{Two simultaneous observations of PSR B1508+55 were performed with the 100-m telescope at Effelsberg and the Five-hundred-meter Aperture Spherical Telescope (FAST). Using and improving existing scintillometry techniques, we leveraged the evolving, very long baseline to precisely measure the screen orientation, effective velocity, and scintillation arc curvature. We inferred the one-screen and two-screen model parameters and we imaged the closer screen.}
{Each single epoch leads to much tighter angular constraints than long-term monitoring of scintillation arcs, revealing an ongoing evolution of the orientation of the closer screen. Images of the scattered pulsar were obtained with a resolution on the order of 0.1 mas. These results confirm the highly anisotropic alignment of the scattered images, while also revealing small-scale deviations from a large-scale straight line.}
{We demonstrate that simultaneous observations of scintillation can be used as a powerful substitute for very long baseline
inferometry.}


\maketitle
\nolinenumbers

\section{Introduction}

Scintillation is an interference pattern manifested over time and frequency. It is a common phenomenon in pulsars whose radiation in the radio regime is often scattered into multiple paths by the free electrons in the ionized interstellar medium (IISM). These paths are spatially coherent if the source is very compact, and hence interfere with each other. This understanding of pulsar scintillation has been accepted since the work by \cite{1968Natur.218..920S}. Nevertheless, the physical nature of the structures causing scintillation remains largely unknown, with competing geometries proposed thus far, such as turbulence \citep[][]{1977ARA&A..15..479R,1985ApJ...288..221C}, magnetotails of clouds \citep{2007ASPC..365..299W}, reconnection sheets \citep{2014MNRAS.442.3338P}, noodle-like filaments \citep{2019MNRAS.486.2809G,2019MNRAS.489.3692G}, or discrete scatterers \citep{2024MNRAS.531.3950K}. Moreover, the connection to larger-scale associates that are observable with other methods than scintillation is unclear. Practically all known objects in the IISM have been hypothesized, whereby evidence has been collected for \ion{H}{ii} regions \citep[e.g.,][]{2022MNRAS.511.1104M}, supernova remnants \citep[e.g.,][]{1973ApJ...181..875R}, and bow shocks in pulsar wind nebulae \citep[e.g.,][]{2025NatAs...9.1053R}. In addition, associations with \ion{H}{i} filaments have been found by \citet{2025ApJ...980...80S}. However, the role of other structures such as the local bubble \citep[e.g.,][]{2024MNRAS.527.7568O} is still unclear and many sources of scattering have no known associates. 

Scintillometry is the science of using scintillation to infer properties of astrophysical objects. It is very sensitive to tiny structures because scintillation preserves some phase information of the incoming radiation, enabling an approach akin to interferometry. Scintillometry has been transformed with the introduction of the Fourier transform of the dynamic spectrum as a standard tool, which was named the secondary spectrum by \citet{1997MNRAS.287..739R}. Secondary spectra have revealed the scintillation arc phenomenon as a new scintillation observable, which was first noted by \citet{2001ApJ...549L..97S} to be ubiquitous in pulsar scintillation. Their parabolic structure was derived by \citet{2004MNRAS.354...43W} and \citet{2006ApJ...637..346C} as a result of scattering by a thin plasma screen. \citet{2003ApJ...599..457H,2005ApJ...619L.171H} first described the remarkable stability of these arcs and their substructure over frequency and time, which can be directly mapped onto the distribution of propagation paths or images. The distribution of these images often seems to be very anisotropic and can even form a straight line of a large number of separated images. This was observed by \citet{2010ApJ...708..232B}, who first combined scintillometry and very long baseline interferometry (VLBI). 

While most scintillometry studies focused on single-dish observations, methods to combine it with VLBI to get superior and instantaneous astrometry of scattering screens have been improved recently: \citet{2019MNRAS.488.4952S} developed a method to make use of single-dish intensities in addition to visibilities and \citet{2023MNRAS.525..211B} combined VLBI with the $\theta$-$\theta$ transform. Both methods were demonstrated on the observations of PSR B0834+06 by \citet{2010ApJ...708..232B}. Applications to other observations are at the point of writing limited to PSR B1508+55 by \citet{2021MNRAS.506.5160M} and PSR B1133+16 by \citet{2025ApJ...992..192S}.

The pulsar B1508+55 is a bright and isolated pulsar discovered by \citet{1968Natur.219..576H} with a very high proper motion and is located at a distance of $2.10^{+0.13}_{-0.14}\,\text{kpc,}$ as measured by \citet{2009ApJ...698..250C}. Prior to late 2020, its scintillation arcs did not show arclets, but flat stripes \citep{2007ASPC..365..254S,2021MNRAS.506.5160M,2022ApJ...941...34S}. Its current scintillation features prominent inverted arclets, which makes it a promising target for another multitelescope study. The transition to the new state happened during observations by \citet{2022MNRAS.515.6198S}, who interpreted the stripes as an effect of a second screen. The first screen lies at a distance within the first 10\% from the Earth to the pulsar while the second one lies in the last 10\%. The presence of the farther screen can still be observed in the new state of scintillation, even though the second arc ended up disappearing. The characteristic stripe-like blurring of secondary spectra as well as the temporal modulation of the amplitude of individual features along the scintillation arc remained visible in 2023 and 2024 at the time of the new observations presented here. However, the transition seems to arise solely from changes in the screen closer to Earth. It is correlated in time with the line-of-sight crossing some bright scattered images of the pulsar as observed at much lower frequencies by \citet{2018evn..confE..17W}. Using the annual variation of scintillation arcs over several years, the movement of tracked features along the arcs, and the consistent modulation by the second screen, a model of the location, velocity, and orientation of anisotropy of the scattering screens was constructed by \citet{2022MNRAS.515.6198S}. 

The aim of this work is to test whether two telescopes observing scintillation simultaneously can measure scattering screens with comparable or even superior sensitivity compared to long-time scintillation arc monitoring, without having to rely on the measurement of visibilities and additional telescopes. This is not only important as a proof of concept, but it would also deliver instantaneous 2D maps of the screen that cannot be obtained from long-time monitoring with a single dish.

This paper is organized as follows. The observations are described in \cref{Sec:Data} and the theoretical background is given in \cref{sec:theory}. The cross-spectra and images of the scattering screen are reported in \cref{sec:imaging}, while the analysis methods used to constrain scattering screen models are presented in \cref{sec:measurement}. The implications are discussed in \cref{sec:conclusion}.

\section{Observations and data processing}
\label{Sec:Data}

\begin{figure}
    \centering
    \includegraphics[trim={0cm 1.5cm 0cm 1.5cm},clip,width=\linewidth]{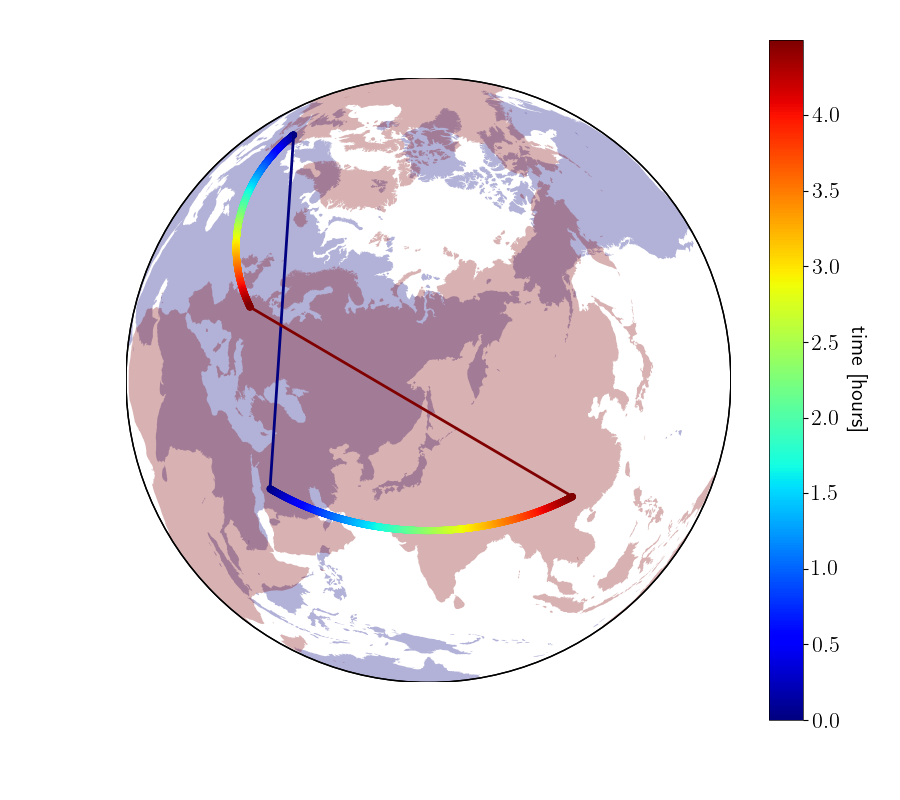}
    \caption{Telescope positions as viewed from the pulsar throughout the observations.  Color represents the time since the start of the observation. Position of the land masses  at the beginning (blue) and  end (red) of the observations and that of the baseline between the telescopes are also shown.}
    \label{fig:Earth}
\end{figure}

Observations of PSR B1508+55 were made simultaneously with the Five-hundred-meter Aperture Spherical Telescope (FAST) and the Effelsberg 100-m telescope. To cover a range of different velocity vectors of Earth and to observe a possible screen evolution, three observations were spread throughout the course of one year. One observation failed due to technical problems at Effelsberg. The two successful observations were performed between December 4 and 5 in 2023 and on May 20 in 2024. The pulsar was observed for the full 4.5 hours of its elevation being high enough to be observed at FAST. The modified Julian dates (MJD) at the time of highest elevation at FAST were 60283 and 60450. The baselines covered are shown in \cref{fig:Earth}. The positions of the telescopes and the declination of the pulsar allow for a wide range of baseline orientations during a single observation.

To maintain flexibility in the channelization, baseband data were recorded at Effelsberg using the Effelsberg Direct Digitisation (EDD) system \citep{EDD}. To cross-correlate the data, we aimed for both datasets to have the same bin widths in terms of time and frequency. Furthermore, it is benficial to have the same edge coordinates of these bins. These goals are challenging to achieve because of the different bandwidth of the receivers as well as the properties of DSPSR \citep{2011PASA...28....1V} which was used for folding and channelization, introducing small variations in the start time and temporal bin width. While folding at the period of the pulsar as given by the ephemeris obtained from the ATNF pulsar catalogue \citep{2005AJ....129.1993M}, the data were divided into subintegrations of the same manually fixed duration at both telescopes. For this number, approximately seven rotations of the pulsar were chosen, whose period is $\sim$\SI{0.74}{\second}. This number is a balance between the temporal resolution and signal-to-noise ratio (S/N). The observed band at FAST with the 19-beam receiver ranged over 500\,MHz from 1000 to 1500\,MHz and was divided into 8192 channels. At Effelsberg, the P217mm receiver was used, which ranges over 400 MHz from 1200 to 1600 MHz, although not the whole band is usable. The different bandwidth makes it impossible to match the FAST channels with an integer number of channels of same width. Thus, the data were divided into 32768 channels, which were five times finer than the FAST channels. The time bins were aligned as well as possible with the \verb|Lepoch| option of DSPSR that lets the first time bin start at a given time corresponding to the first time bin of the FAST data. 

The functions \verb|dedisperse()|, \verb|remove_baseline()|, and \verb|convert_state('Intensity')| of the Python wrapper of PSRCHIVE \citep{2012AR&T....9..237V} were used to remove the baseline, convert the data to Stokes I, and to dedisperse them. At this point, the data consisted of data cubes over time, frequency, and rotational phase separated into 64 phase bins.

Radio frequency interference (RFI) is an important limitation for this method because of its demand for high S/N within short subintegrations and small channels at both telescopes. Using a custom implementation in Python, strong known RFI bands at Effelsberg were manually masked while the remaining data were cleaned by investigating statistics in the manually identified off-pulse region of each subintegration and channel. The investigated statistics were median, mean, modulation index (standard deviation divided by mean), kurtosis, skewness, range (difference between maximum and minimum), maximal Fourier coefficient divided by mean, and Pearson cross-correlation coefficient between neighboring phase bins. A function, $P$, predicting the likelihood of a sample not being RFI based on its vector, $\bm{x}$, of off-pulse statistics was calibrated on manually labeled samples. We chose a multivariate Gaussian for this function and fit for the mean and standard deviation of each statistic by minimizing
\begin{equation}
    \left\langle P(\bm{x}) \right\rangle_{\text{RFI}} + \left\langle 1 - P(\bm{x}) \right\rangle_{\text{not RFI}}. \, 
\end{equation}
This fit converged after collecting $\sim 100$ of randomly chosen samples of each category, which can be done in short time by using intermediate fits to search for candidates of the category of lower sample number. The statistic with the most prediction power for RFI turned out to be the range followed with some distance by the kurtosis. Channels where more than $75\%$ of subintegrations had a likelihood of less than 0.5 to not be RFI were masked for the whole observation. All completely masked data were treated as if their $P$ was zero.

Next, template pulse profiles were formed by averaging over all subintegrations, subtracting the median of the off-pulse region for each channel, and taking the mean over all channels, in this order. Masked values were ignored. Finally, all off-pulse phase bins of the obtained profile were set to zero, along with any remaining values smaller than zero at the edges of the pulse.

The dynamic spectrum was then formed from data that had not been masked in advance. Also, the median of the off-pulse region was subtracted for each subintegration and channel. Afterward, the data were multiplied by the template profile and then summed over all phase bins. The described method was used to dynamically remove background and noise level, while measuring the flux of the pulsar while taking into account the whole pulse, with phases of high S/N being given more weight.

All dynamic spectra were now converted to the same frequency channels that were chosen to be equal to the channels of the FAST data, but extended to the band covered by Effelsberg. This was done by binning the channels of the Effelsberg data into these common frequency grids, where the average was weighted with $P^2$. The remaining RFI (often caused by contamination solely in the on-pulse region) was suppressed by decreasing the weights of data values larger than seven times the weighted standard deviation of all data and masking all data values greater than ten times above it. These thresholds were determined from past observations of this pulsar in \citet{2022MNRAS.515.6198S} that showed that the real signal almost never exceeds them. Even so, such contamination could not be completely removed. In addition, the square root of the mean weight of each bin was used to obtain the $P$ value for the whole bin. In the converted data, all bins with $P<0.5$ were masked. In summary, almost 50\% of data needed to be masked in both Effelsberg observations and around 10\% of the data in both FAST observations. Hence, only data from 1300\,MHz to 1425\,MHz were used. In this subband, both telescopes were sensitive enough to observe with less than 3\% of the data masked as RFI.

\begin{figure*}
    \centering
    \includegraphics[trim={0cm 1cm 0cm 0cm},clip,width=\linewidth]{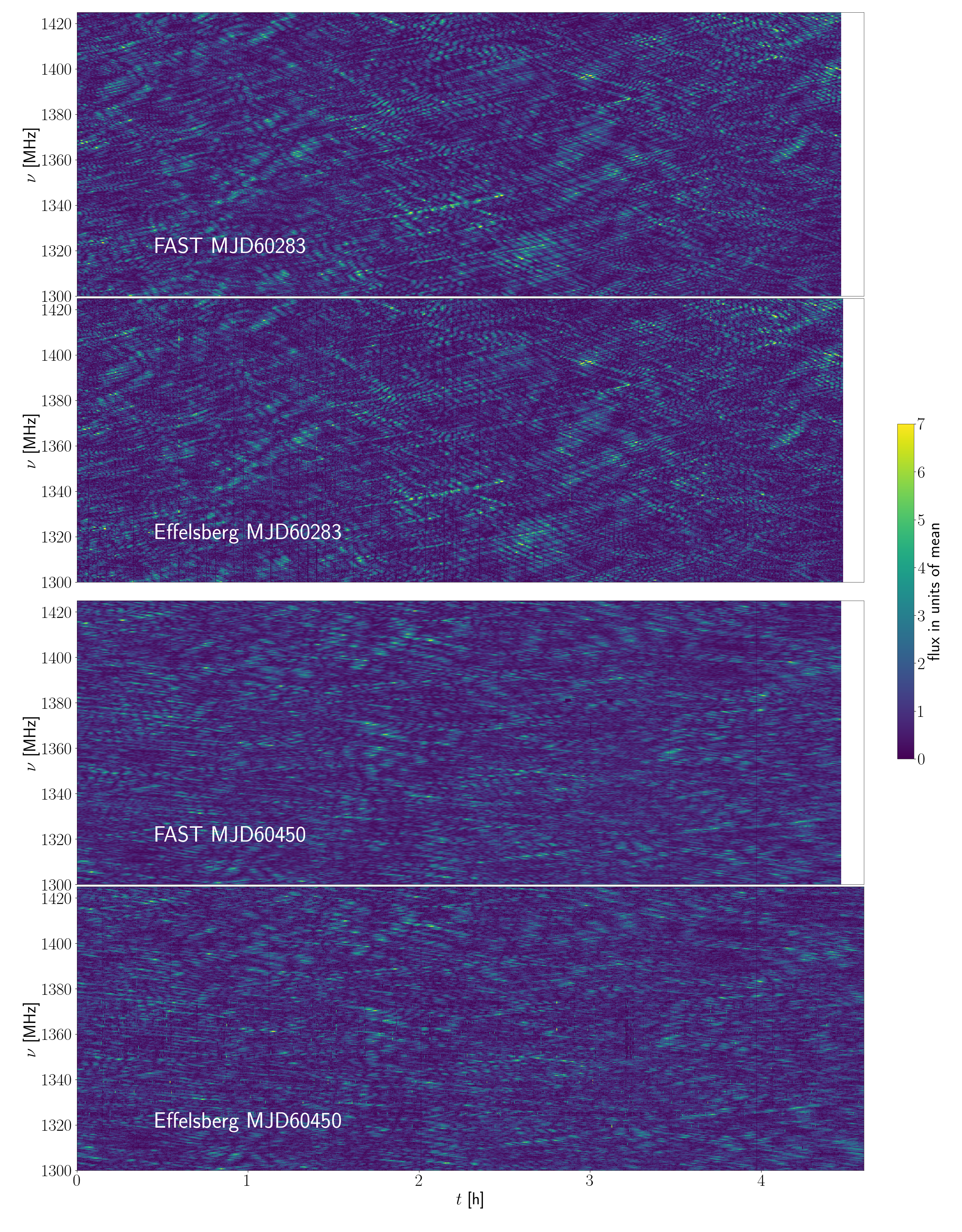}
    \caption{Dynamic spectra of the two simultaneous observations. The data have been normalized to a mean of unity in every time bin and the band has been cut to the optimal subband. The time axis is synchronous and measured from the beginning of simultaneous observations.}
    \label{fig:DynSpecs}
\end{figure*}

Finally, the dynamic spectra were normalized to remove intrinsic pulse-to-pulse variations and remaining frequency-dependent variations caused by the bandpass applied at the telescopes. This was done iteratively. First, each time bin was divided by its mean over frequency, not counting the masked data. Here, time bins whose mean was below 10\% of the average of the whole dynamic spectrum were also masked because these would contaminate the subsequent analysis with their low S/N. Then, the same normalization was done for every channel by dividing by the mean over all time bins and it was repeated for every time bin to ensure that the mean of each time bin is exactly unity. The resulting dynamic spectra are shown in \cref{fig:DynSpecs}.

Using \textsc{Astropy} \citep{2022ApJ...935..167A}, the telescope positions corresponding to the observation times are obtained in the Geocentric Celestial Reference System (GCRS), while the Earth's velocity is obtained in the International Celestial Reference System (ICRS). These were then projected to 2D vectors in right ascension (RA) and declination (Dec) using the position of the pulsar.

\section{Scintillation arcs from multiple telescopes}
\label{sec:theory}

PSR B1508+55's scintillation contains dominant 1D scattering and has shown strong signs of two scattering screens as reported in \citet{2021MNRAS.506.5160M} and \citet{2022MNRAS.515.6198S}. In 2020, it  underwent a phenomenological change, which also affected the annual variation of its scintillation arcs that is indicative of a change in orientation of the axis of anisotropy of the scattering medium \citep{2022MNRAS.515.6198S} within the closer screen. Thus, we did not treat this angle as a constant in our modeling.

We applied both a one-screen and a two-screen model here. Although past studies have strongly implied a second screen, this screen's properties are much less constrained; whereas a one-screen model strengthens the comparability to other studies that apply similar models. Furthermore, the second screen is likely to be much farther away and hence smaller in angular size. Thus, a one-screen model is expected to be able to describe any signal arising from the difference between two observer locations on Earth to a good approximation.

\begin{figure}
    \centering
    \includegraphics[width=\linewidth]{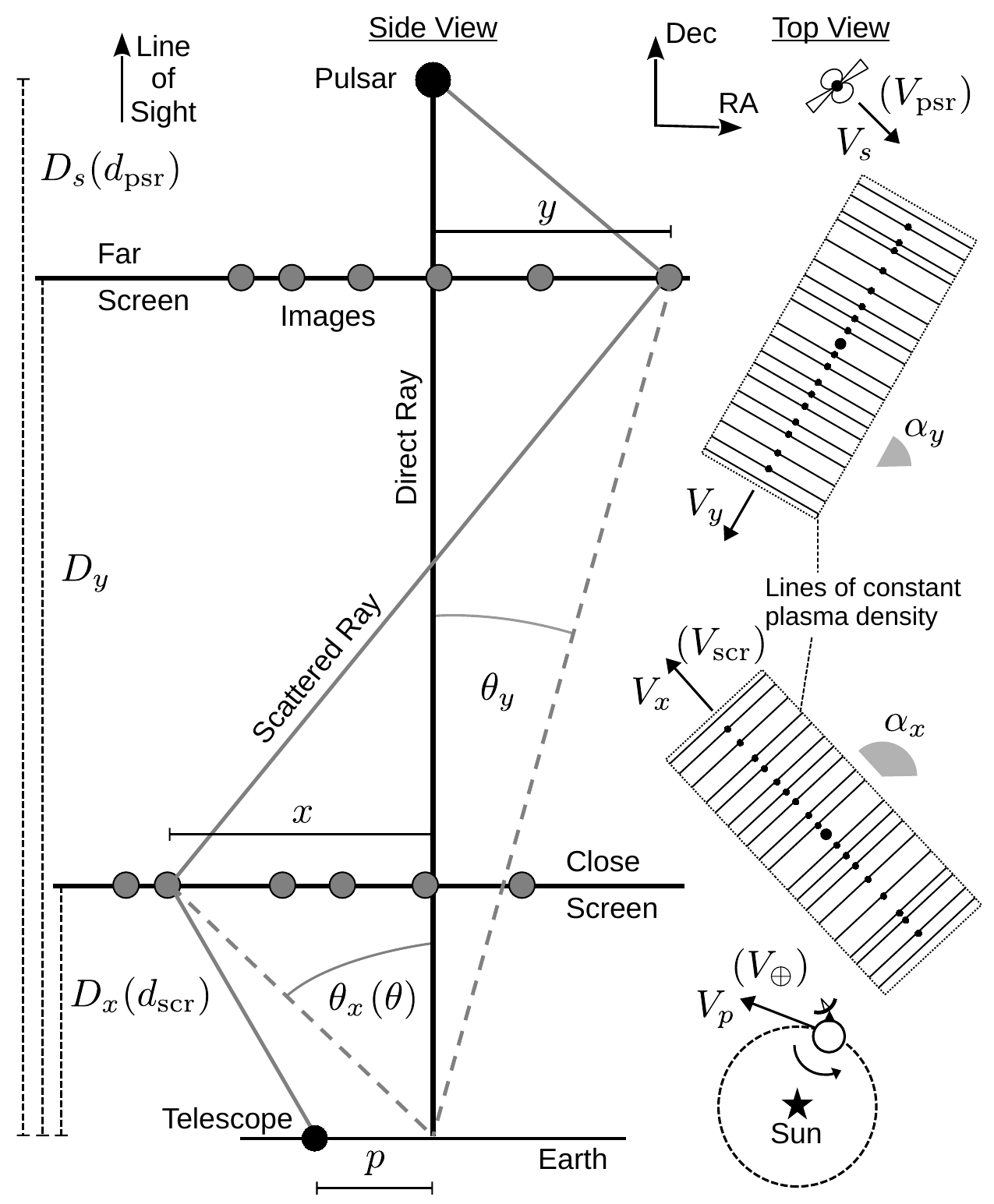}
    \caption{Definitions relevant to this study. The directions of the arrows are examples and do not define the positive direction of signed quantities. Alternative notations in the one-screen case are given in parentheses.}
    \label{fig:Sketch_Definitions}
\end{figure}

All symbols used for geometric variables are defined in \cref{fig:Sketch_Definitions}. Scintillation is insensitive to small variations in positions along the line of sight. Thus, all positions and velocities can be treated as 2D vectors. The coordinates are defined along the positive directions of RA and Dec. The angles $\alpha_x$ and $\alpha_y$ of the orientation of straight lines of scattered images are defined such that they are zero in the direction of positive RA and grow toward the direction of positive Dec. This notation is the same as used in \citet{2022MNRAS.515.6198S} and can be converted to angles measured east of north via $\alpha_\text{East of North}=90^\circ-\alpha$.

\subsection{One-screen theory}

The intrinsic temporal and spectral variation of the pulsar is removed by the normalization. The remaining variations are dominated by interference of different paths of propagation whose phase differs due to the different path lengths. This geometric phase difference $\Phi$ is simply given by the number of wavelengths, $\lambda=c/\nu$, that fit into the length difference, $\Delta L$, of the scattered path of propagation and the pulsar's distance along the direct line of sight, such that $\Phi = 2\pi \Delta L/\lambda$. In the small-angle approximation, this length depends quadratically on the scattering angle.  Since at least the work by \citet{2010ApJ...708..232B}, it has been customary to express $\Delta L = D_\text{eff} \bm{\theta}^2/2$, using the effective distance
\begin{equation}
    D_\text{eff} = \frac{d_\text{scr} d_\text{psr}}{d_\text{psr}-d_\text{scr}}
,\end{equation}
where $d_\text{scr}$ is the distance from the observer to the screen and $d_\text{psr}$ is the distance from the observer to the pulsar. Temporal evolution happens because of changes of the observed scattering angle, $\bm{\theta}$, due to the relative velocities of telescopes, the ISM, and the pulsar. An effective velocity term can be defined as
\begin{equation}
    \bm{V}_\text{eff} = \bm{V}_\oplus - \frac{d_\text{psr}}{d_\text{psr}-d_\text{scr}}\bm{V}_\text{scr} + \frac{d_\text{scr}}{d_\text{psr}-d_\text{scr}}\bm{V}_\text{psr}
,\end{equation}
such that the angular velocity of locations fixed on the screen is $\der \bm{\theta}/\der t=-\bm{V}_\text{eff}/D_\text{eff}$. This definition of the effective velocity is not unique in the current literature. It is a convenient definition in the case of multiple telescopes because it is equal to the velocity of an observer with respect to a constant point in the scintillation pattern. Hence, the offset, $\bm{p}$, of a telescope from the center of the Earth can simply be added to the cumulative shift by this velocity to obtain the phase of a scattered ray exactly at the position, $\bm{p}(t),$ of a telescope in the plane perpendicular to the line of sight at a point, $t$, in time,
\begin{equation}
    \Phi = \frac{\pi\nu }{c}D_\text{eff}\,\left(\bm{\theta} -\frac{\bm{p}(t)+\bm{V}_{\text{eff}}t}{D_\text{eff}}\right)^2,
\end{equation}
where we have chosen to define $\bm{V}_\oplus$ as the velocity of the center of the Earth while any movement of the observer on its surface relative to the center is encoded in $\bm{p}(t)$. All angles are defined with respect to the line of sight to the pulsar at $t=0$ which is the beginning of the observation. In the case of a 1D screen, this expression can be shortened because only contributions that depend on the angular position $\theta$ along one axis of the screen vary between rays. Removing all constant contributions and keeping only the components of vectors that are parallel to the straight line of images, we obtain
\begin{equation}
    \Phi = \frac{2\pi\nu}{c}\left[ \frac{D_\text{eff}}{2}\theta^2-\left(p_\shortparallel(t)+V_{\text{eff},\shortparallel}t\right)\,\theta \right]. \,
\end{equation}
The baseline parallel to the screen between two telescopes $A$ and $B$ can be defined as 
\begin{equation}
\bm{b}(t_A,t_B) = \bm{p}_{A}(t_A)-\bm{p}_{B}(t_B) \, . \label{eq:def_baseline}
\end{equation}
It can be easily read off now that telescope $A$ will observe exactly the same situation as telescope $B$ after a time, $\Delta t$, that is given by
\begin{equation}
    \Delta_t (t_A,t_B) = -b_{\shortparallel}(t_A,t_B) / V_{\text{eff},\shortparallel} \, . \label{Eq:time_shift}
\end{equation}
The times, $t_A$ and $t_B$, at which the data were taken do not need to be equal. The parallel baseline is given by
\begin{equation}
    b_{\shortparallel}(t_A,t_B) = \bm{b}(t_A,t_B)\cdot\begin{pmatrix}\cos \alpha_x \\ \sin \alpha_x \end{pmatrix} \, .
\end{equation}

Pairs of individual scattered paths can be made visible by forming secondary spectra, which are the square modulus of the 2D Fourier transform of the intensity dynamic spectrum, where the Doppler rate, $f_\text{D}$, and delay, $\tau$, are the Fourier conjugates of time and frequency, respectively. \citet{2004MNRAS.354...43W} and \citet{2006ApJ...637..346C} first derived that secondary spectra represent a map of the scattering screen in differential coordinates, where
\begin{align}
    f_\text{D}(\bm{\theta}_1,\bm{\theta}_2,t,\nu) &= \frac{\der}{\der t}\,\frac{ \Phi(\bm{\theta}_1,t,\nu)-\Phi(\bm{\theta}_2,t,\nu) }{2\pi} \, , \\
    \tau(\bm{\theta}_1,\bm{\theta}_2,t,\nu) &= \frac{\der}{\der \nu}\,\frac{ \Phi(\bm{\theta}_1,t,\nu)-\Phi(\bm{\theta}_2,t,\nu) }{2\pi}, \ 
\end{align}
for all pairs $(\bm{\theta}_1,\bm{\theta}_2)$ of scattering angles. Even for fixed image positions $\bm{\theta}_1$ and $\bm{\theta}_2$, the corresponding delay and Doppler rate evolve with time and frequency. Since secondary spectra cannot be formed from a single data point, this results in a smearing effect for which different mitigation techniques have been proposed by \citet{2014JGRA..11910544F} and \citet{2021MNRAS.500.1114S}. The secondary spectrum coordinates of the 1D one-screen case are
\begin{align}
    f_\text{D} &= -\frac{\nu}{c} V_{\text{eff},\shortparallel} \, \left(\theta_1-\theta_2\right) -\frac{\nu}{c}\frac{\der}{\der t}p_\shortparallel(t)\, \left(\theta_1-\theta_2\right), \,  \label{eq:fD_1s1d}\\
    \tau &= \frac{D_\text{eff}}{2c}\left(\theta_1^2-\theta_2^2\right) - \frac{p_\shortparallel(t)+V_{\text{eff},\shortparallel}t}{c}(\theta_1-\theta_2). \, \label{eq:tau_1s1d}
\end{align}
Usually, the second terms in \cref{eq:fD_1s1d,eq:tau_1s1d} are neglected. They cause a periodic variation of the scintillation arc over the course of a day. We confirmed that our scintillation arc measurements are not sensitive enough to identify such a variation and, hence, we chose to neglect these terms in the following. A double-parabolic appearance of scintillation arcs and inverted arclets that both follow $\tau=\eta f_\text{D}^2$ can be read off from these relations. Thus, secondary spectra containing such features are strong evidence for 1D scattering, as is the case for PSR B1508+55. Adopting the approach by \citet{2022MNRAS.515.6198S}, we reformulated the arc curvature, $\eta$, in terms of the effective drift rate, $\zeta$, to avoid frequency dependencies and divergences,
\begin{equation}
    \zeta = \frac{1}{2\nu\sqrt{\eta}} = \frac{\vert V_{\text{eff},\shortparallel} \vert}{\sqrt{2c D_\text{eff}}} \, . \label{eq:arc_curvature}
\end{equation}

\citet{2019MNRAS.488.4952S} developed generalizations for the secondary spectrum in the case of multistation observations. The intensity cross-spectrum, $S_I$, can be formed without having to measure the visibilities and is defined as
\begin{equation}
S_I(f_\text{D},\tau) = \tilde{I}_A(f_\text{D},\tau) \times \tilde{I}_B(-f_\text{D},-\tau)
,\end{equation}
where $\tilde{I}$ is the Fourier transform of the dynamic spectrum, $I$. The complex phase of $S_I$ is given by
\begin{equation}
    \Phi\left(f_\text{D},\tau\right) = -\frac{2\pi\nu}{c}\bm{b}\cdot(\bm{\theta}_1-\bm{\theta}_2) \overset{\text{1D}}{=} -\frac{2\pi\nu}{c}b_\shortparallel(\theta_1-\theta_2) \, .
    \label{Eq:cross_phase}
\end{equation}

The intensity cross-spectrum is the Fourier transform of the cross-correlation function, just as the secondary spectrum is the Fourier transform of the autocorrelation function. For 1D screens, the cross-correlation function is equal to the autocorrelation function shifted by $\Delta_t$, defined in \cref{Eq:time_shift},
\begin{equation}
    \text{CCF}(\Delta t) = \text{ACF}(\Delta t-\Delta_t)
\end{equation}
where $\Delta t$ is the time lag for which the correlation is computed. This is a result of the scintillation being identical at the two telescopes. This situation arises necessarily for a 1D screen but can only be observed in 2D screens for the unlikely case of the effective velocity and the baseline being exactly aligned.

The temporal ACF of scintillation can be modeled as a Gaussian and is defined by the scintillation time, $t_s$,
\begin{equation}
    \text{ACF}(\Delta t) = m^2\,\exp\left(-\frac{\Delta t^2}{2 t_s^2}\right) \, .
    \label{eq:gaussian_ACF}
\end{equation}
Here, $m$ is the modulation index which is equal to the standard deviation of the data divided by its mean. An exponential including the exponent $5/3$ instead of 2 has been proposed as an alternative to a Gaussian to fit the temporal ACF \citep[e.g.,][]{2019MNRAS.485.4389R}. However, testing a different function is not necessary here because only the peak of the model function is relevant for a measurement of the shift, $\Delta_t$.

\subsection{Two-screen theory}

An analytical solution for the wave phase in a model of two 1D screens, where each ray is scattered by both screens has been developed in \citet{2022MNRAS.515.6198S} and written down in Eqs. A8 and A15 therein. Instead of setting $p=0$, we are  able to obtain effective positions, $p_\text{eff}$, that are different for each screen,
\begin{equation}
\begin{split}
    \Phi =& \frac{\pi \nu}{c}\Big( D_\text{eff,x}\theta_x^2 + D_\text{eff,y}\theta_y^2 -2 D_\text{mix} \theta_x \theta_y\\
 &-2(p_{\text{eff},x} + V_{\text{eff},x,\shortparallel}t)\theta_x -2(p_{\text{eff},y} + V_{\text{eff},y,\shortparallel}t)\theta_y \Big) \, .
\end{split}
\label{eq:2screen_phase}
\end{equation}
The distance and velocity terms are given in \citet{2022MNRAS.515.6198S}. The effective positions result from the different projections of the telescope position on the two 1D screens and can be obtained with the same logic. They are given by
\begin{align}
    p_{\text{eff},x} &= p_{\shortparallel} +  \frac{ \gamma \delta D_x D_{y,s} p_{\perp}}{D_y D_{x,s} - D_x D_{y,s}\delta^2} \, ,\\
    p_{\text{eff},y} &= - \frac{ \gamma D_y D_{x,s} p_{\perp}}{D_y D_{x,s} - D_x D_{y,s}\delta^2} \, ,
\end{align}
where $\gamma = \sin(\alpha_x-\alpha_y)$ and $\delta = \cos(\alpha_x-\alpha_y)$.

Each screen produces a different time shift that is obtained by dividing the effective baseline by the effective velocity of that screen, analogously to \cref{Eq:time_shift}. These time shifts could be observed separately and are not added up, such that the dynamic spectrum at the two telescopes becomes more different than just being shifted. In particular, the temporal shift between two telescopes $A$ and $B$ due to the first screen is given by
\begin{equation}
    \Delta_t (t_A,t_B) = -\frac{p_\text{eff,x,A}(t_A)-p_\text{eff,x,B}(t_B)}{V_{\text{eff},x,\shortparallel}}  \, . \label{Eq:time_shift_2scr}
\end{equation}

\section{Imaging the scattered pulsar}
\label{sec:imaging}

The secondary spectrum is closely connected to the scattered image of the pulsar. If it is sufficiently sparse, each point in it represents a single pair of subimages. Its power is their multiplied intensity and its location is determined by their relative phase. As such, the secondary spectrum is a holograph \citep{2008MNRAS.388.1214W}, preserving phase information within the intensity data. The idea behind the work by \citet{2010ApJ...708..232B} was to combine this phase information with the phase information obtained via VLBI. \citet{2010ApJ...708..232B} and \citet{2019MNRAS.488.4952S} used the fact that at the apex of each arclet one of the two interfering images lies at the origin to create images of the scattering screen of PSR B0834+06. In summary, the location along the scintillation arc separates subimages from each other whose VLBI phase can then be used to infer their angular positions.

Without the VLBI information, the pulsar can be imaged from the secondary spectrum information alone. One example for doing so is given in \citet{2021MNRAS.500.1114S}. However, very sharp arclets are required. The orientation of the obtained image on the sky is not constrained and the locations are ambiguous with two solutions on each side of the effective velocity vector. Since the arclets of PSR B1508+55 are blurrier than those of PSR B0834+06 and contaminated by the stripe effect, such a single-dish imaging method was not performed.

\subsection{Intensity cross-spectra}

\begin{figure*}
    \centering
    \includegraphics[trim={0cm 1cm 0cm 0cm},clip,width=\linewidth]{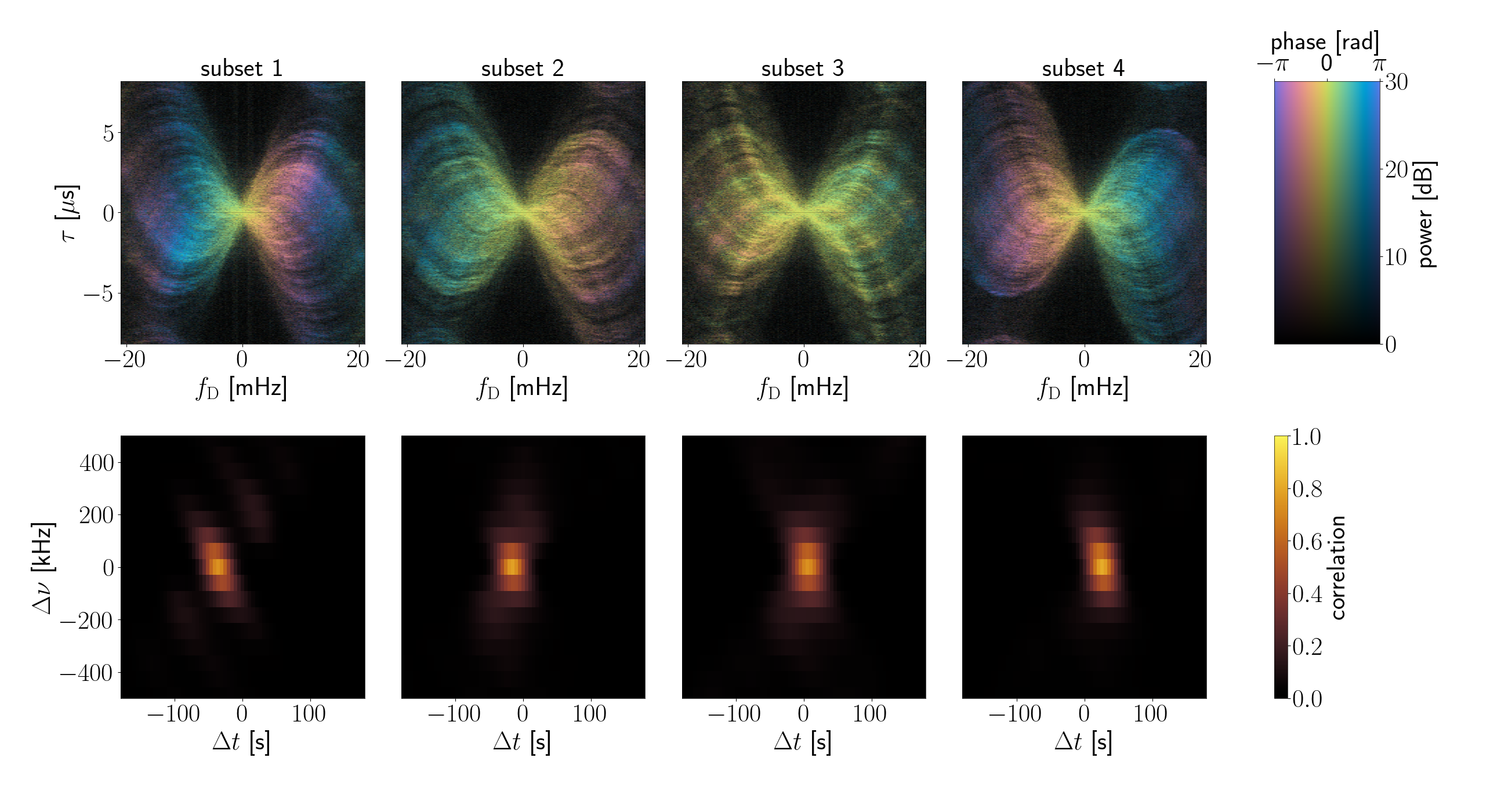}
    \caption{Cross-secondary spectra (top row) and cross-correlations (bottom row) from December 4-5, 2023. The observation has been split into four parts of equal length in time, shown in chronological order. The colors denote the complex phase of the cross-secondary spectra, while their brightness denotes the power. The cross-correlation functions are obtained as the Fourier transform of the cross-secondary spectra. They illustrate the temporal shift corresponding to the linear phase evolution in the cross-spectra.}
    \label{fig:CrossSpectra2023}
\end{figure*}

\begin{figure*}
    \centering
    \includegraphics[trim={0cm 1cm 0cm 0cm},clip,width=\linewidth]{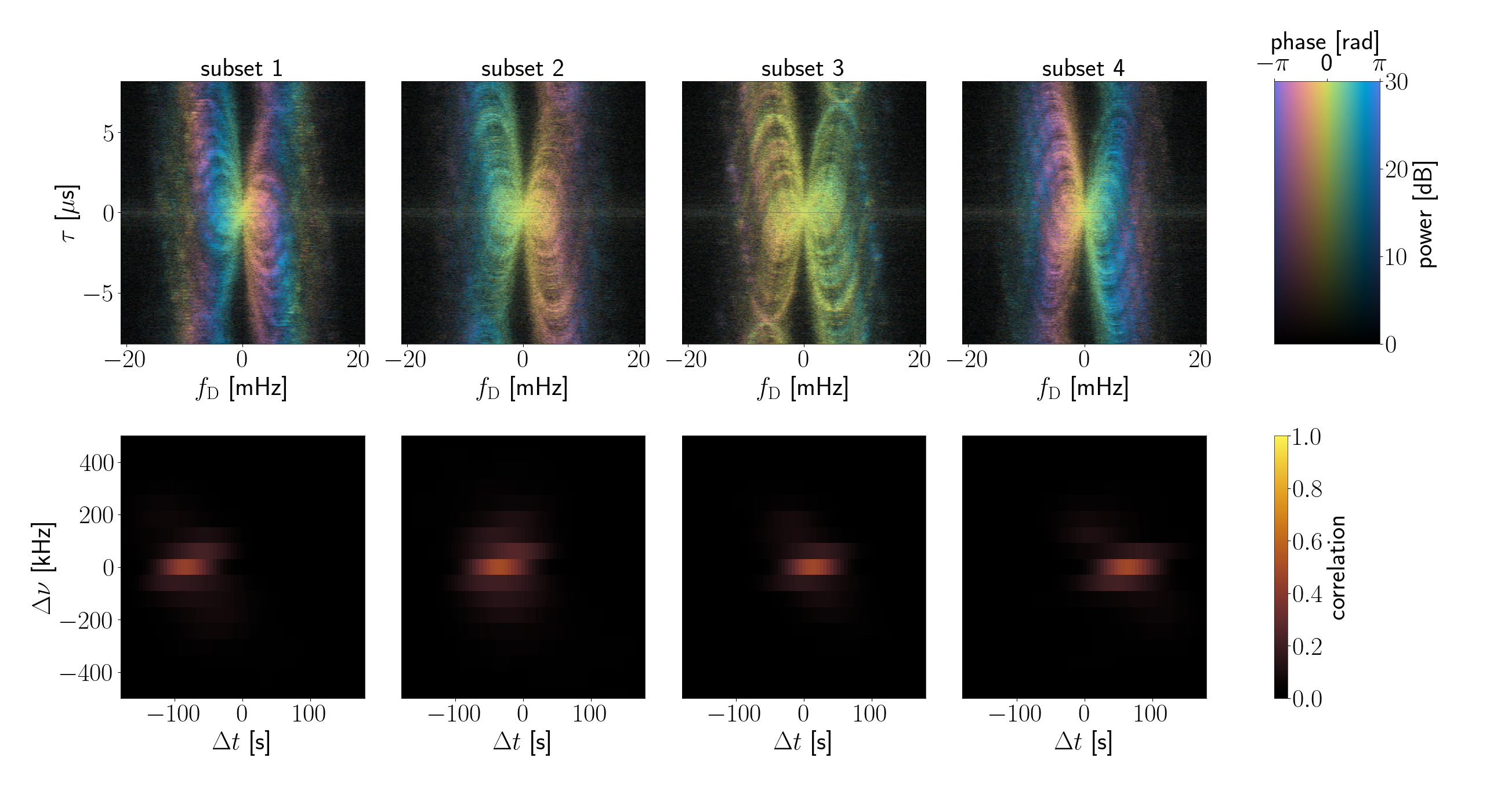}
    \caption{Cross-secondary spectra (top row) and cross-correlations (bottom row) from 20 May 2024. The setup is identical to \cref{fig:CrossSpectra2023}.}
    \label{fig:CrossSpectra2024}
\end{figure*}

To form the intensity cross-spectra, both dynamic spectra need to be located on an identical and regular grid in time and frequency. The procedure to get identical frequency channels and approximately identical time bins was described in \cref{Sec:Data}. In addition, small shifts between the coordinate grids can be corrected by considering the Fourier transform, $\tilde{I}'$, of a dynamic spectrum whose coordinates are shifted by $\delta t$ in time and by $\delta\nu$ in frequency,
\begin{equation}
\begin{split}
    \tilde{I}'(f_\text{D},\tau) &= \int I(t+\delta t,\nu+\delta\nu)\,\text{e}^{-2\pi i (f_\text{D}t+\tau\nu)}\text{d}t\text{d}\nu \\
    &= \text{e}^{2\pi i (f_\text{D}\delta t+\tau\delta\nu)}\,\tilde{I}(f_\text{D},\tau)
\end{split}
.\end{equation}
This factor was divided out before the computation of any cross-spectrum. Evolving shifts between time bins due to corrections of the time of arrival of the pulses could not be corrected with this method.

In contrast to the case discussed in \citet{2010ApJ...708..232B}, the baseline evolved significantly throughout the observation and the phase of the cross-spectrum as given in \cref{Eq:cross_phase} is expected to vary strongly as a result. Thus, meaningful cross-spectra can only be obtained from subsets of the data. In \cref{fig:CrossSpectra2023} and \cref{fig:CrossSpectra2024}, cross-spectra are shown that were computed from subsets each consisting of a fourth of the observation time. Clear linear phase evolutions are visible in these spectra in addition to the double parabolic shape of the scintillation arc.

A phase evolution that is linear in Doppler rate and independent of delay corresponds to a simple temporal shift of the dynamic spectra at the two telescopes with respect to each other. Hence, the corresponding cross-correlation functions that are also reported in \cref{fig:CrossSpectra2023} and \cref{fig:CrossSpectra2024} show the correlation to be maximal at an offset from the origin in time lag.

\subsection{Producing the image}
\label{sec:image}

\begin{figure}
    \centering
    \includegraphics[trim={2cm 8cm 3cm 1cm},clip,width=\linewidth]{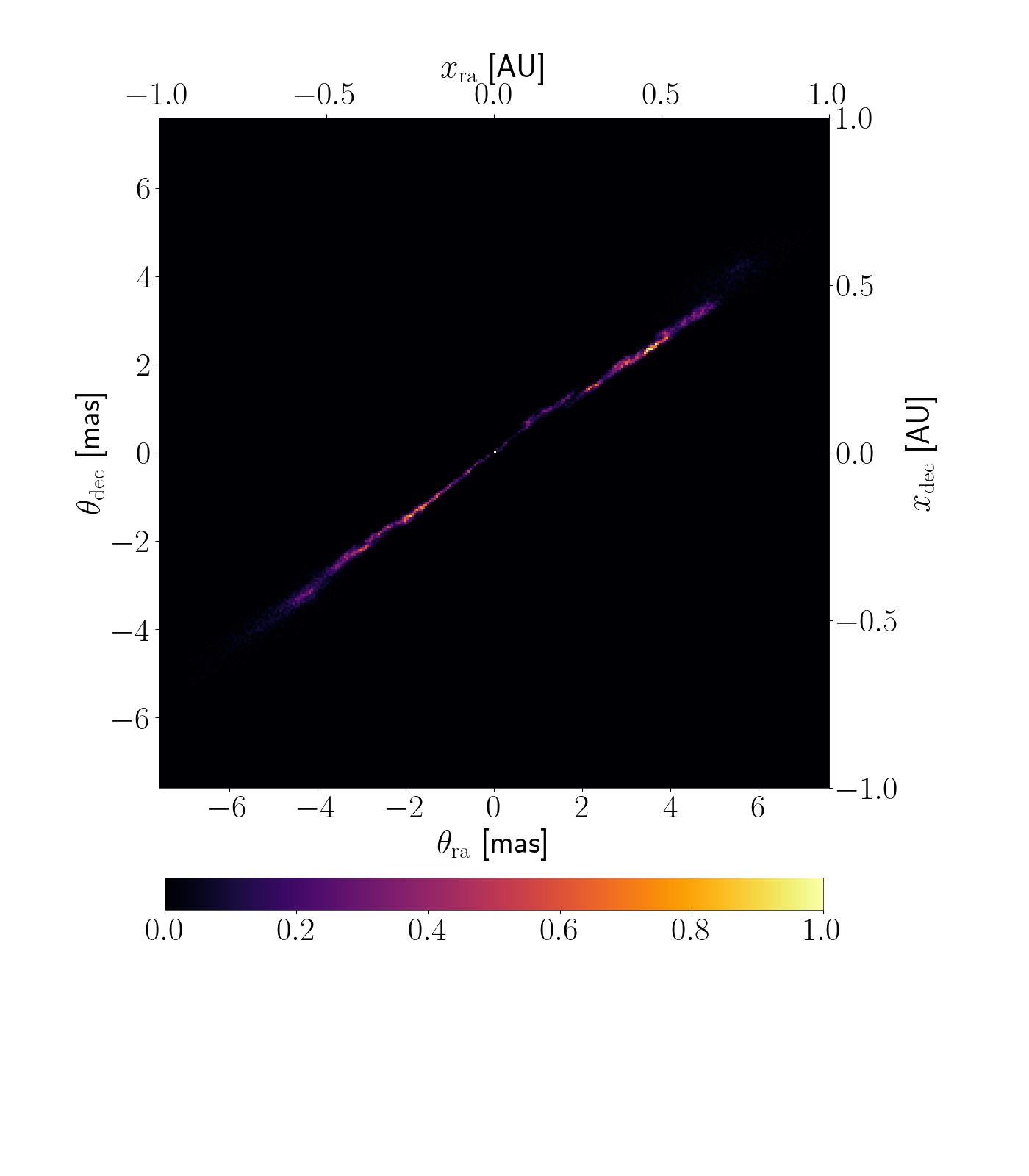}
    \caption{
    Image of the scattering screen at December 4-5, 2023. The central point cannot be imaged and is instead shown as fully saturated to highlight the position of the pulsar. The color scale is linear and normalized to the highest value. A resolution of about \SI{0.1}{mas} is achieved. The evolution of the image over the course of the observation can be found as an online movie.
    }
    \label{fig:Image2023}
\end{figure}

\begin{figure}
    \centering
    \includegraphics[trim={2cm 8cm 3cm 1cm},clip,width=\linewidth]{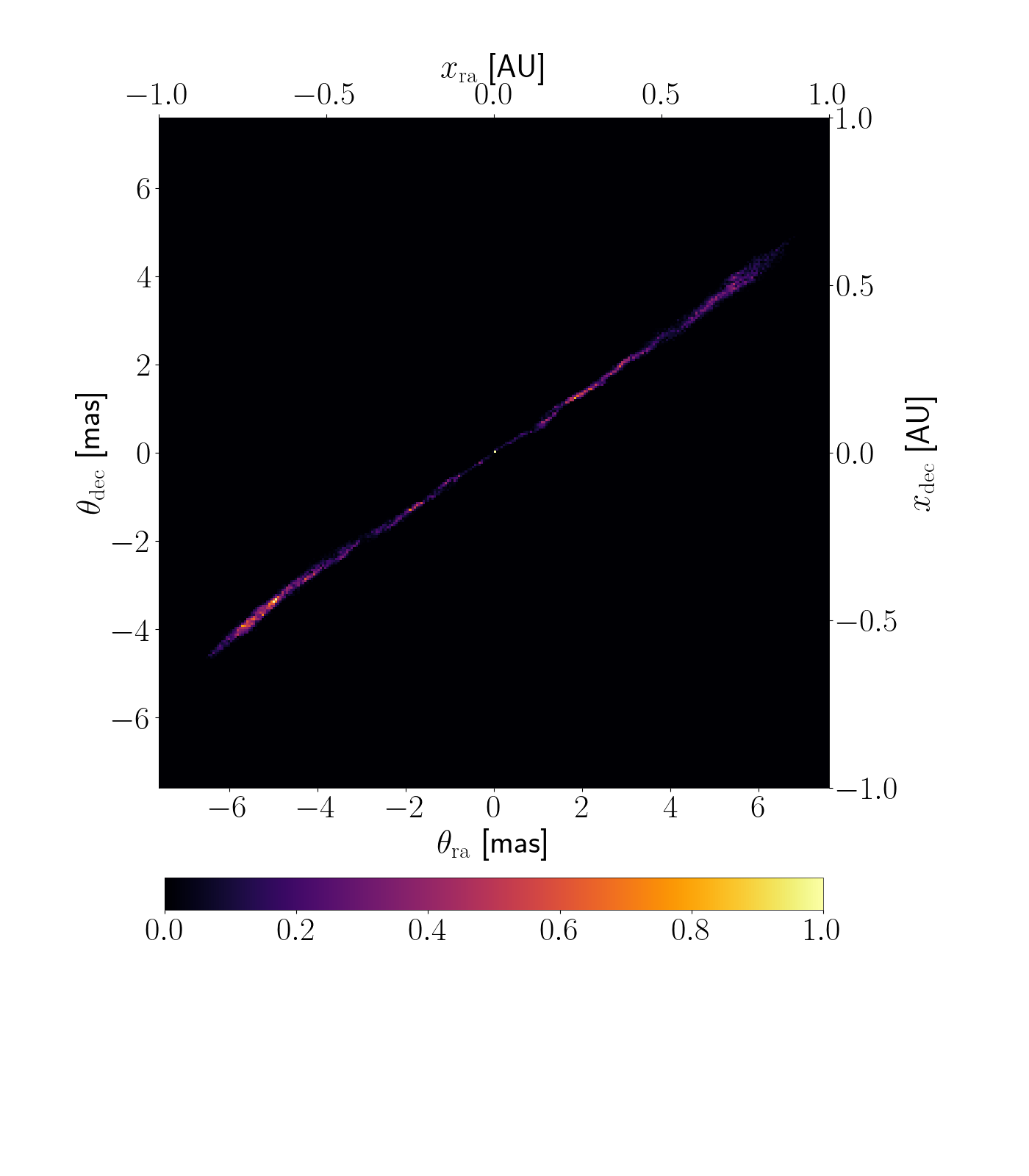}
    \caption{
    Image of the scattering screen at 21 May 2024. See \cref{fig:Image2023} for more details. An online movie is available.
    }
    \label{fig:Image2024}
\end{figure}

The apexes of inverted arclets lie in the vicinity of the main arc. These locations result from pairs of angles where one angle is the origin. According to \cref{Eq:cross_phase}, the phase of the intensity cross-spectrum at these points becomes
\begin{equation}
    \Phi_\text{apex}\left(f_\text{D}(\bm{\theta}),\tau(\bm{\theta})\right) = -\frac{2\pi\nu}{c}\bm{b}\cdot\bm{\theta} \, .
    \label{eq:2D_apex_cross_phase}
\end{equation}
Even though we only have a single pair of telescopes, we can leverage the evolution of this baseline and divide the observation into subsets to obtain multiple baselines. Thus, \cref{eq:2D_apex_cross_phase} can be solved for $\bm{\theta}$ to form an image. We developed an imaging method closely related to the one described by \citet{2019MNRAS.488.4952S}.

First, the dynamic spectra were divided into ten adjacent subsets of equal length in time. To obtain smaller bins in $f_\text{D}$, they were each padded in time with zeros by twice the subset length at each side. The obtained cross-spectra where then smoothed by a top-hat function of \SI{0.1}{\micro\second} in $\tau$ and \SI{0.5}{\milli\hertz} or \SI{0.25}{\milli\hertz} in $f_\text{D}$ where the former was used for the observation in 2023 and the latter for the one in 2024. As a result of the smoothing, the quality of the phases is sufficient to be unambiguously corrected for wrapping at $\pm\pi$.

Next, points are selected which lie close enough to the main arc such that \cref{eq:2D_apex_cross_phase} can be applied. The scintillation arc's coordinates are known from measurements of the curvature that are described in \cref{sec:curvature} below. We estimated by trial and error that the same $f_\text{D}$ windows as those used for the smoothing define a region were our approximations are valid. For each point of positive delay in this region, \cref{eq:2D_apex_cross_phase} can be solved by a least-squares optimization for $\bm{\theta}$, inserting the baselines of the central time of each subset. The solutions were binned into a 300$\times$300 grid spanning over two astronomical units (AU) according to the screen distance at the two epochs taken from the second column of \cref{tab:mcmc_results} below. Each point was weighted by the square root of the mean cross-spectrum amplitude multiplied by $\vert f_\text{D}\vert$, which corresponds to its amplitude in angular space according to \citet{2021MNRAS.500.1114S}. The amplitude at the origin cannot be measured.

The resulting images are shown in \cref{fig:Image2023,fig:Image2024}. This method contains several parameters that were optimized manually and affect the achieved resolution. Still, the dominant structures within these images are stable with respect to some variation of these parameters. These contain the expected straight line but also some substructure deviating from it, as well as brighter regions that relate to individual arclets. The method described here assumes that all images lie close enough to a straight line for their corresponding arclets to be close to the main arc. Offset arclets would need to be imaged separately. We also neglected contributions of the second screen to the phase in this method. The result confirms that they are indeed negligible.

After the solutions for each point are found, the amplitudes in the weighting can be replaced. This can be used to obtain the temporal evolution of the image by using secondary spectra created from a moving window of the dynamic spectrum. It confirms the modulation of the second screen moving along the image that was described in \citet{2022MNRAS.515.6198S}. 

\section{Measurement of screen parameters}
\label{sec:measurement}

The cross-spectrum analysis revealed that the most prominent difference between the scintillation at the two telescopes is an evolving temporal shift. It can also be seen by eye in \cref{fig:DynSpecs}. This shift has been theoretically predicted in \cref{Eq:time_shift} and \cref{Eq:time_shift_2scr} for a one-screen and a two-screen model, respectively. In a one-screen model, measuring this shift between the telescopes while the baseline rotates with the Earth provides a direct measure of the screen orientation and the parallel effective velocity. At the same time, the arc curvature in \cref{eq:arc_curvature} is given by the parallel effective velocity and the effective distance. Thus, measuring the arc curvature in combination with the varying interferometric shift unambiguously constrains the parameters $\alpha_x$, $D_\text{eff}$, and $V_{\text{eff},\shortparallel}$. We used these measurements to constrain one-screen and two-screen models.

\subsection{Arc curvature measurement}
\label{sec:curvature}

Despite being the defining parameter of a scintillation arc, the arc curvature can be difficult to obtain precisely. Even under ideal observational conditions, the signal itself may significantly depart from the ideal double-parabolic structure because of deviations from a 1D alignment of images. The images might even evolve in time and frequency, be further scattered by additional screens, or follow a distribution not centered at the origin due to large-scale refraction.

Multiple methods have been proposed to improve or automate the measurement of arc curvatures. \citet{2016ApJ...818...86B} and \citet{2020ApJ...904..104R} introduced methods based on the Hough transform where the power along different trial arcs is summed and the best curvature is obtained by maximizing this sum. This method works better the more concentrated arcs are. For the opposite case of an arc that is rather double-parabolic rather than a single parabola, \citet{2022MNRAS.510.4573B} developed a method that explicitly uses the 1D nature of the screen corresponding to this case. This is the method adopted here and we refer to it as the $\theta$-$\theta$ method. Its idea is to use the $\theta$-$\theta$ transform introduced in \citet{2021MNRAS.500.1114S} that transforms arclets into straight lines for different trial curvatures. For the correct curvature, these lines are horizontal and vertical. This condition is equivalent to maximizing the highest eigenvalue of an eigenvector decomposition of the $\theta$-$\theta$ diagram.

\begin{figure*}
\centering
\includegraphics[trim={2cm 0cm 2cm 0cm},clip,width=\linewidth]{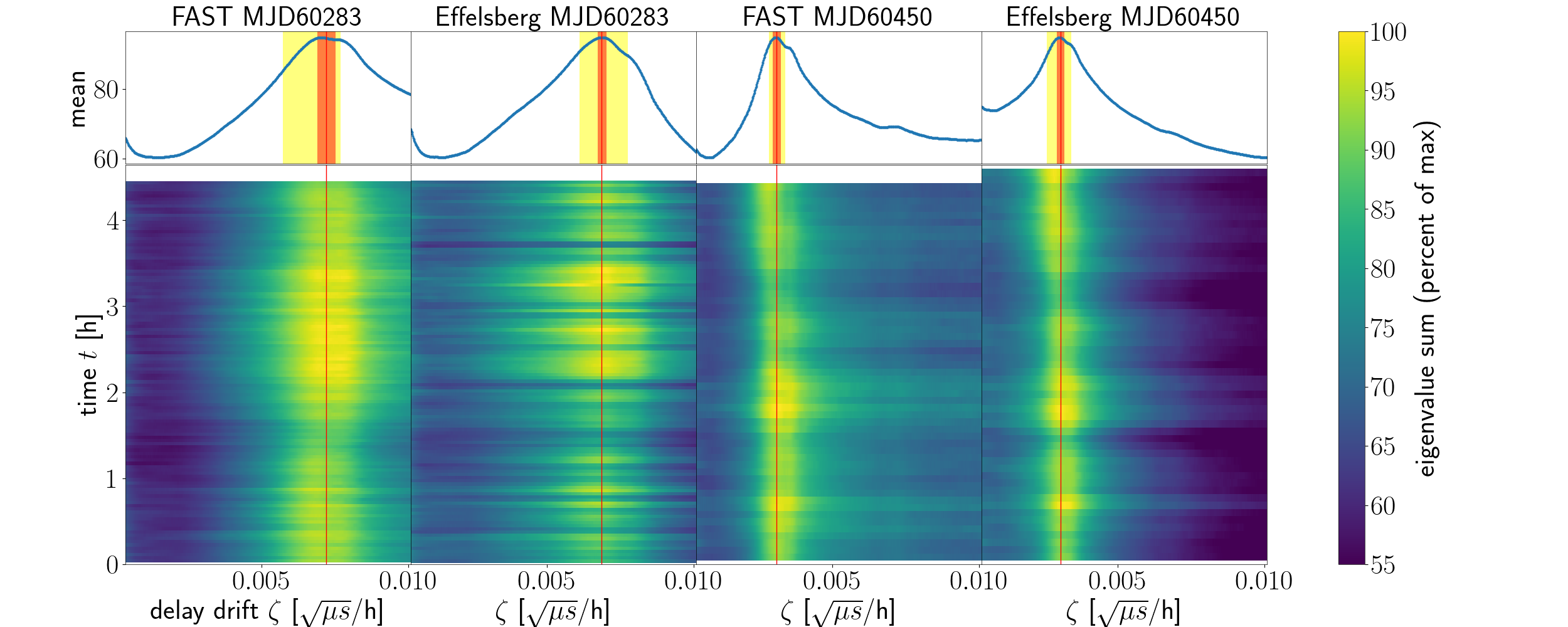}
\includegraphics[trim={2cm 0cm 2cm 0cm},clip,width=\linewidth]{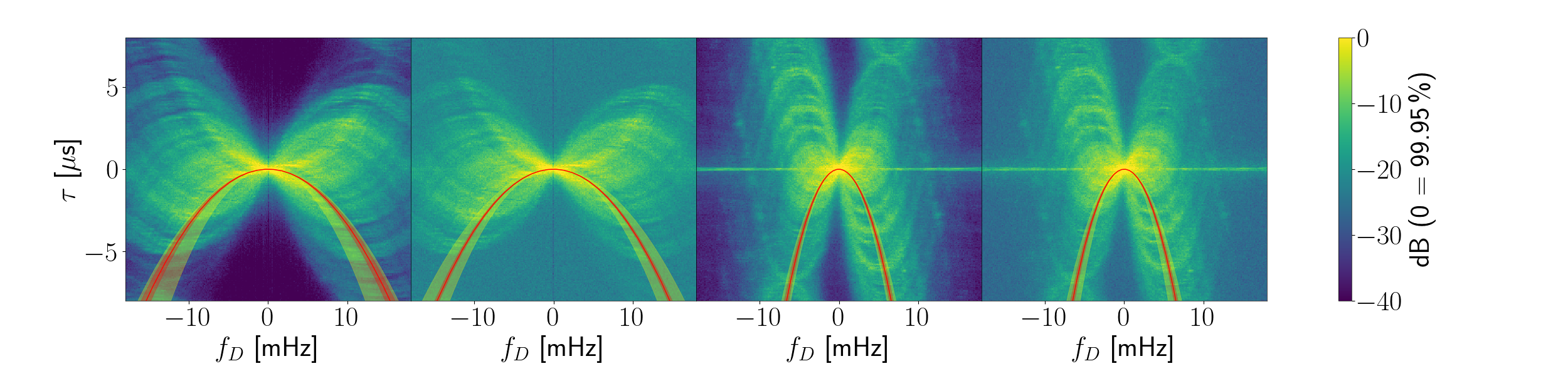}
\caption{
Arc curvature measurement. Bottom row: Secondary spectra with equal color scale gauged by the 99.95 percentile. Mid row: Sum of eigenvalues for all subsets in frequency as a function of subsets in time and trial $\zeta$. To row: Corresponding mean over all subsets in time. The best-fit value for $\zeta$ or the scintillation arc is shown in red. In the top and bottom plots its 1$\sigma$ region is shown in red and the range of possible values identified by eye is shown in yellow as a cross-check.
}
\label{fig:zetas}
\end{figure*}

As discussed in \citet{2022MNRAS.515.6198S}, $\theta$-$\theta$ method is sometimes contaminated by additional peaks that are likely nonphysical. There, a manual method of measurement has been used in addition to cross-check the measured values. As can be seen in \cref{fig:zetas}, the $\theta$-$\theta$ method was also affected here but their effect never shifts the best fit to or over the edges of the manually identified possible range of arc curvatures. Thus, the same methodology as in \citet{2022MNRAS.515.6198S} was used without the combination with manual measurements; namely,~the dynamic spectrum was divided into subsets of size $\Delta t$ in time and $\Delta \nu$ in frequency that were set to
\begin{equation}
\begin{split}
    \Delta t &= 60\,\text{s}\times\frac{0.006\,\sqrt{\mu\text{s}}/\text{h}}{\zeta_\text{est.}}, \, \\
    \Delta \nu &= 2.5\,\text{MHz. \, }
\end{split}
\end{equation}
The value for $\zeta_\text{est.}$ used here was estimated by eye using the graphical method described in \citet{2022MNRAS.515.6198S}. For each of these subsets, $\theta$-$\theta$ transformations were performed for 600 trial values of $\zeta$ that ranged from 0.00036 to 0.01008\,$\sqrt{\mu\text{s}}/\text{h}$. For each of these, the highest eigenvalue of an eigenvector decomposition was saved. These eigenvalues were then summed over all frequencies to reduce scatter and a parabolic fit was performed on a region of width 0.0006\,$\sqrt{\mu\text{s}}/\text{h}$ centered around the maximum value. The mean and standard deviation over all fit results was then used as the measured value and its uncertainty, respectively. The measured values are shown in \cref{tab:zetas}. For subsequent parameter inference, the combined value and standard deviation was computed. The measurements are in agreement, but still greater than the values predicted from the best-fit model by \citet{2022MNRAS.515.6198S}.

\begin{table}
        \centering
    \caption{Arc curvature measurements}
    \label{tab:zetas}
    \begingroup
    \renewcommand{\arraystretch}{1.2}
        \begin{tabular}{lll}
     \toprule
     Telescope & $\zeta$ [$10^{-3}\sqrt{\mu s}$/h] & $\eta_{1400}$ [$\mu$s/mHz$^2$] \\
     \toprule
     \multicolumn{3}{c}{\it  December 4-5, 2023} \\
     \hline
     Effelsberg & $6.86\pm0.13$ & $0.0351^{+0.0014}_{-0.0013}$ \\
     FAST & $7.20\pm0.29$ & $0.0319^{+0.0027}_{-0.0024}$ \\
     combined & $6.92\pm0.12$ & $0.0345^{+0.012}_{-0.012}$ \\
     2022 predictions & $6.592$ & $0.03804$ \\
     \hline
     \multicolumn{3}{c}{\it May 20, 2024} \\
     \hline
     Effelsberg & $3.05\pm0.12$ & $0.178^{+0.015}_{-0.013}$ \\
     FAST & $3.09\pm0.13$ & $0.173^{+0.015}_{-0.013}$ \\
     combined & $3.068\pm0.087$ & $0.176^{+0.010}_{-0.010}$ \\
     2022 predictions & $2.970$ & $0.1874$ \\
     \bottomrule
        \end{tabular}
    \endgroup
\end{table}

\subsection{Correlation analysis}
\label{sec:correlation}

Since the baseline is continuously evolving, correlating the temporal scintillation is problematic. Instead, we cross-correlated the single time bins. This method is sufficient if there is enough structure in the spectral scintillation to differentiate time bins from each other. It has several advantages. First, assumptions on a regular grid in time are no longer required, which removes a potential source of systematic errors. More importantly, it completely removes the correlation peak at zero shift that arises from remaining intrinsic pulse-to-pulse variations because the temporal intensity information is dismissed. This effect would have tended to bias the measured shift toward smaller absolute values. In addition, baselines from telescope positions evaluated at different times can now be correctly taken into account. In summary, the cross-correlation function (CCF) is computed as
\begin{equation}
    \text{CCF}(t_\text{Eff},t_\text{FAST}) = \frac{1}{N_\nu}\sum_n I_\text{Eff}(t_\text{Eff},\nu_n) \,I_\text{FAST}(t_\text{FAST},\nu_n)
,\end{equation}
where $N_\nu$ is the number of channels with center frequencies, $\nu_n$, and the intensities, $I$, were normalized as described in \cref{Sec:Data} and their mean was subtracted.

\begin{figure}
    \centering
    \includegraphics[trim={0cm 1cm 1cm 1cm},clip,width=\linewidth]{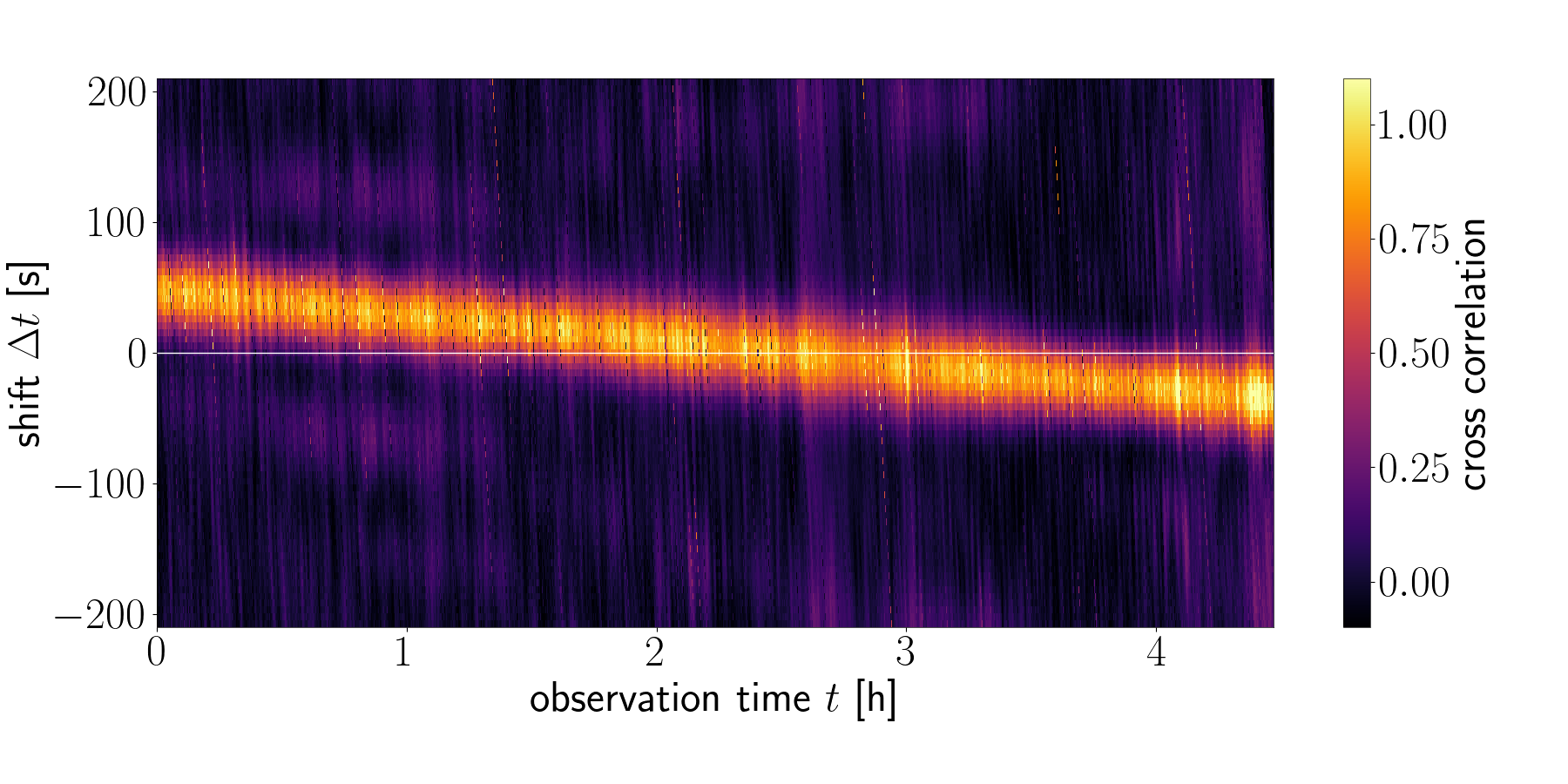}
    \includegraphics[trim={0cm 1cm 1cm 1cm},clip,width=\linewidth]{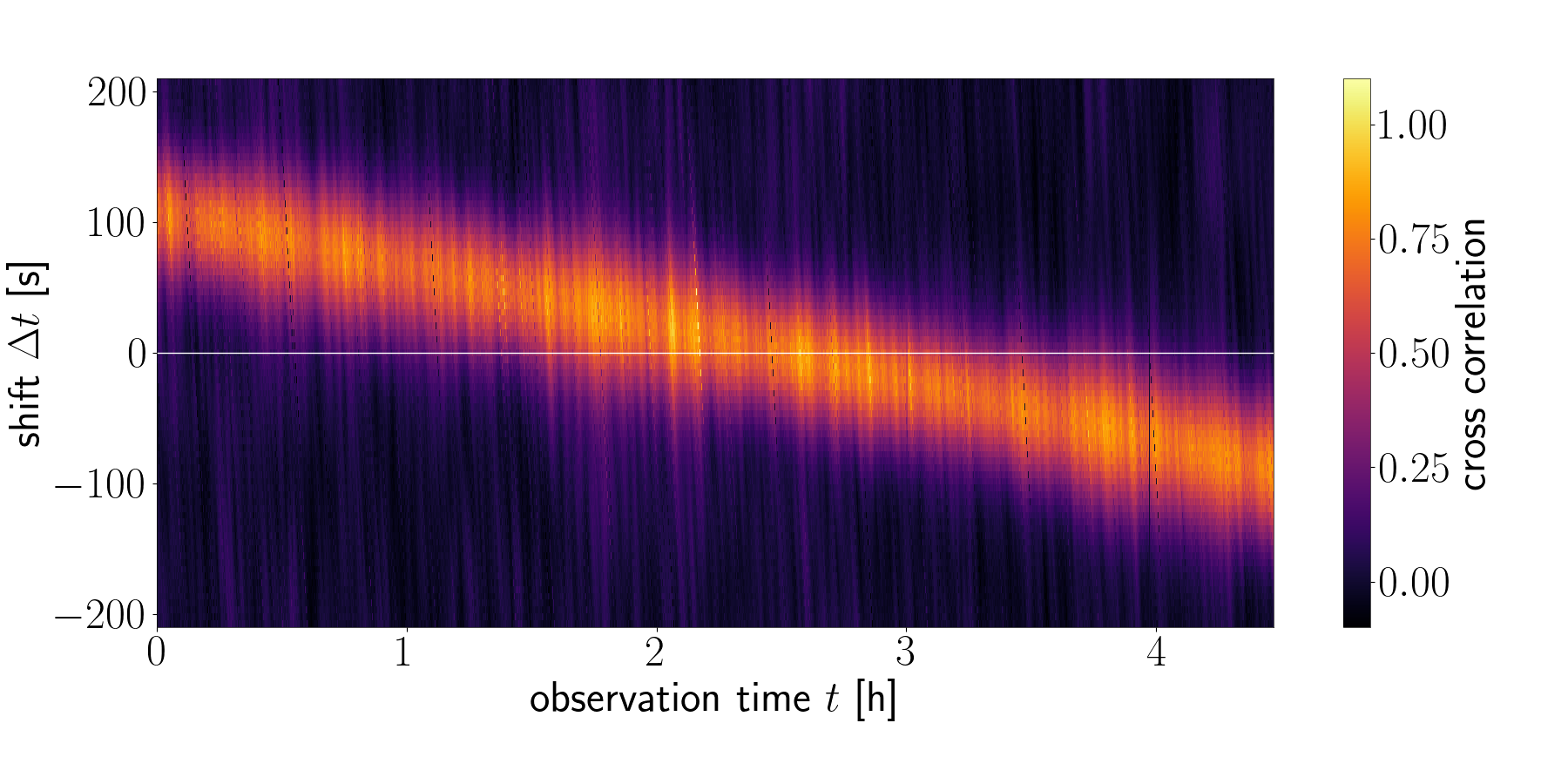}
    \caption{Evolution of the cross-correlation at December 4-5, 2023 (top) and from May 20, 2024 (bottom). Assuming a regular grid, the corresponding time shift between the telescopes is shown. Positive values mean that the scintles arrive at FAST first. The horizontal white line marks vanishing shift between the scintillation.
    }
    \label{fig:CrossCorrelation_shift}
\end{figure}

In \cref{fig:CrossCorrelation_shift}, we show the cross-correlation of each time bin with each other translated to a temporal shift as a function of observation time. This relation can be well described by a linear function with two free parameters. To investigate more complicated fit functions would require a solid uncertainty estimation of the CCF, which has not yet been accomplished because of strong correlations that depend on the amplitude of the signal. Since nearby data points are not independent, the cross-correlation of shifted (but otherwise identical) data are nonzero, even at lags that deviate from the true temporal shift between the two telescopes. Fortunately, the amount of data is large enough to allow for the creation of subsets. Thus, the uncertainty of the two fitted parameters can be estimated as follows. A fit to the full dataset produces values for the two parameters describing the temporal shift as well as values for the modulation index, $m$, and the scintillation time, $t_s$, as defined in \cref{eq:gaussian_ACF}. Then, the CCF was separated into subsets, where the only time bins at Effelsberg that were considered were those separated by $4 t_s$. The fit was repeated on these subsets and the standard deviation of the results for each fit parameter was taken as an uncertainty estimate.

\begin{table}
        \centering
    \caption{Comparison of cross-correlation and autocorrelation analysis of the temporal shift (see \cref{sec:correlation}).}
    \label{tab:correlation_results}
    \begingroup
    \small
        \begin{tabular}{>{\raggedright\arraybackslash}m{1.5cm}m{1.3cm}m{1.4cm}m{1.4cm}m{1.3cm}}
     \toprule
     Dataset & $\alpha_x$ [$^\circ$] & $V_{\text{eff},\shortparallel}$ [km/s] &  $t_s$ [s] & $m$ \\
     \toprule 
     \multicolumn{5}{c}{\it 4-5 December 2023} \\
     \hline
     CCF & 35.877(51) & -97.19(69) & 21.99(12) & 0.9347(64) \\
     ACF Eff. & \multicolumn{2}{c}{\it neglected} & 21.91(17) & 0.9375(86) \\
     ACF Eff. & \multicolumn{2}{c}{\it fixed to CCF values} & 21.95(17) & 0.9376(86) \\
     ACF FAST & \multicolumn{2}{c}{\it neglected} & 21.893(20) & 0.9573(16) \\
     ACF FAST & \multicolumn{2}{c}{\it fixed to CCF values} & 21.820(20) & 0.9573(16) \\
     \hline
     \multicolumn{5}{c}{\it 20 May 2024} \\
     \hline
     CCF & 34.04(10) & -41.69(17) & 39.88(13) & 0.8490(32) \\
     ACF Eff. & \multicolumn{2}{c}{\it neglected} & 39.90(15) & 0.8572(36) \\
     ACF Eff. & \multicolumn{2}{c}{\it fixed to CCF values} & 40.06(15) & 0.8572(36) \\
     ACF FAST & \multicolumn{2}{c}{\it neglected} & 39.390(78) & 0.8794(16) \\
     ACF FAST & \multicolumn{2}{c}{\it fixed to CCF values} & 39.078(77) & 0.8795(16) \\
     \bottomrule
        \end{tabular}
    \endgroup
\end{table}

For a one-screen model, a natural choice for the two fit parameters are the effective velocity, $V_{\text{eff},\shortparallel}$, along the screen and the orientation, $\alpha_x$, of the screen. Then, the temporal shift is given by \cref{Eq:time_shift}. Results for this parametrization are shown in \cref{tab:correlation_results}. The two parameters can be independently constrained because the orientation sets the time of zero shift and the effective velocity sets the slope of the time shift. For a two-screen model, the parameters obtained by the fit need to be translated according to \cref{Eq:time_shift_2scr} as
\begin{align}
    \alpha &\mapsto \arctan\left( \frac{\sin(\alpha_x)+\beta\cos(\alpha_x)}{\cos(\alpha_x) - \beta\sin(\alpha_x)} \right) ,\\
    V_{\text{eff},\shortparallel} &\mapsto \frac{ V_{\text{eff},x,\shortparallel}}{\sqrt{1+\beta^2}},
\end{align}
where
\begin{equation}
    \beta = \frac{ \gamma \delta D_x D_{y,s}}{D_y D_{x,s} - D_x D_{y,s}\delta^2} \, .
\end{equation}

For comparison, the same analysis can be repeated by correlating the dataset  from each telescope to itself instead, which yields the autocorrelation function (ACF). Then, the modulation index and scintillation time scale of the expected Gaussian ACF defined in \cref{eq:gaussian_ACF} can directly be fitted without any shift. However, this means that the evolution of the telescope's position has been neglected. Different positions on Earth view the scintillation pattern at different times, which affects the observation time at which a certain value of autocorrelation is to be expected. This effect can be accounted for by computing the expected shift along the baseline between the positions of the telescope at the two different times correlated, just as the shift due to the baseline given by \cref{eq:def_baseline} between two different telescopes was computed. Since the ACF drops to zero quickly, these single-station baselines are too small for $\alpha_x$ and $V_{\text{eff},\shortparallel}$ to be independently measured. Instead, we fix them to the values obtained fitting to the CCF. 

\cref{tab:correlation_results} shows the results for the four fit parameters for the CCF and for the ACF;  for the latter, the results are shown both in the case where we are accounting for the additional effect of single-telescope baselines as well as neglecting it. The difference between both is statistically significant, even though the absolute difference is tiny. The absolute values of $m$ are quite similar reflecting the visible similarity of the observations. The more sensitive FAST data are in statistical tension with both the CCF and the ACF results from Effelsberg. This can also be explained by contributions of data gaps, RFI, and noise because small differences can also be seen in the inferred scintillation time. Thus, the result allows only for small differences of the signal arriving at the two telescopes, confirming the dominant one-dimensionality of the screen. \citet{2021MNRAS.506.5160M} reported the opposite result of a low cross-correlation between simultaneous observations with a very long baseline. However, the corresponding observations took place before the mentioned transition in 2020, when the farther screen strongly affected the dynamic spectra, which means that the results are not directly comparable.
For each simultaneous observation, the sensitivity on the screen orientations is at least four times higher than that obtained from long-time monitoring in \citet{2022MNRAS.515.6198S}.

\subsection{Model fitting}
\label{sec:fitting}

\begin{table}
        \centering
    \caption{Model best fits and MCMC standard deviations.}
    \label{tab:mcmc_results}
        \begin{tabular}{lll}
     \toprule
      & All data, two screens & New data, one screen \\
     \toprule
     $\alpha_{x,2024}$ \hfill [$^\circ$] & $34.00\pm0.10$ & $34.04\pm0.10$ \\
     $\alpha_{x,2023}$ \hfill [$^\circ$] & $35.85\pm0.05$ & $35.88\pm0.05$ \\
     $\alpha_{x,2020-22}$ \hfill [$^\circ$] & $37.42\pm0.41$ & - \\
     $\alpha_{x,\text{WS}}$ \hfill [$^\circ$] & $54.21\pm2.38$ & - \\
     \midrule
     $D_{x,2024}$ \hfill [pc] & \multirow{2}{*}{$131.7\pm0.9$} & $121.8\pm6.6$ \\
     $D_{x,2023}$ \hfill [pc] &  & $129.6\pm4.6$ \\
     \midrule
     $V_{x,\shortparallel,2024}$ \hfill [km/s] & $3.6\pm0.5$ & $6.8\pm3.2$ \\
     $V_{x,\shortparallel,2023}$ \hfill [km/s] & $3.4\pm0.4$ & $6.6\pm2.1$ \\
     \bottomrule
        \end{tabular}
\end{table}

To infer the screen's orientation, distance, and velocity, we performed a least-squares fit and a Markov chain Monte Carlo (MCMC) analysis using \textsc{emcee} \citep{2013PASP..125..306F}. This was done on two different combinations of data and model. The results are shown in \cref{tab:mcmc_results}. The available data consist of the arc curvatures, screen orientations, and effective velocities at the two epochs of simultaneous observations, as listed in \cref{tab:zetas} and \cref{tab:correlation_results}, where the results from the CCF analysis were used. In addition, we used all the data that had previously been used for the parameter inference in \citet{2022MNRAS.515.6198S}, which consists of Effelsberg observations from 2019 to 2022. To differentiate, we refer to the two simultaneous epochs as the "new data," while the combination of both sources is labeled "all data". The new data were incorporated into the updated likelihood function as independent Gaussian random variables. 

First, a two-screen model was fit to all the data. As described in \cref{sec:imaging}, the screen orientation evolved over time. Hence, we introduce the two new parameters $\alpha_{x,2023}$ and $\alpha_{x,2024}$ for the screen orientation at the two simultaneous epochs in addition to the parameters defined in \citet{2022MNRAS.515.6198S}. The projected screen velocity, $V_{x,\shortparallel}$, was derived from the absolute value and orientation that were used as fit parameters. The obtained distance, $D_x$, of the closer screen is much larger than the previous constraint of $127.1\pm1.5\,$pc. This deviation is partially offset in the multidimensional space of parameters with the $\chi^2/\text{ndf}$ increasing from 0.95 to 1.1. Nevertheless, this finding points toward the assumption of a constant screen distance possibly being wrong. A shift of \SI{\sim 5}{pc} over three years where the line of sight only travels \SI{\sim 0.0001}{pc} along the screen is rather unlikely, which makes a nonzero thickness of the screen a more physical alternative than an inclined thin screen. The screen's velocity is still in agreement with a constant value. The orientation, $\alpha_x$, of the screen is evolving in a consistent direction and remains incompatible with being constant in a two-screen model.

\begin{figure}
    \centering
    \includegraphics[trim={0cm 0.5cm 11.1cm 11.3cm},clip,width=.49\linewidth]{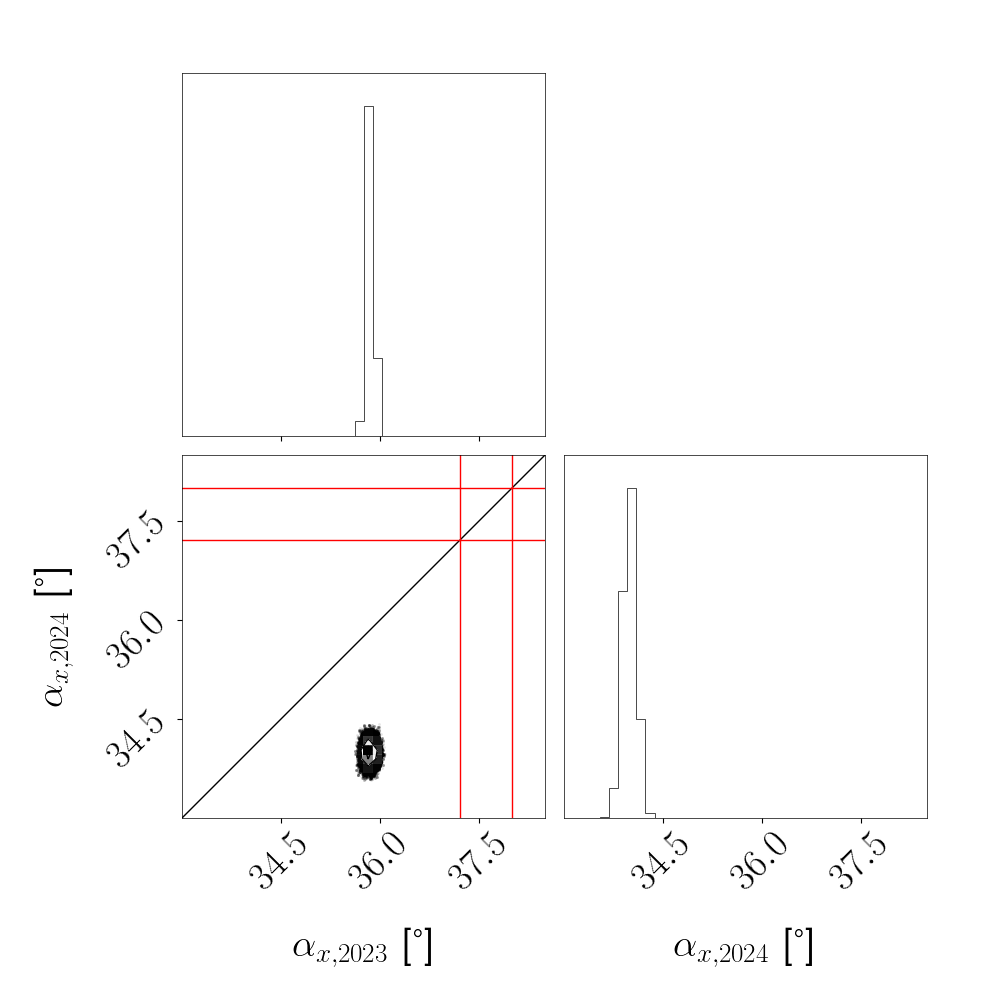}
    \includegraphics[trim={0cm 0.5cm 11.1cm 11.3cm},clip,width=.49\linewidth]{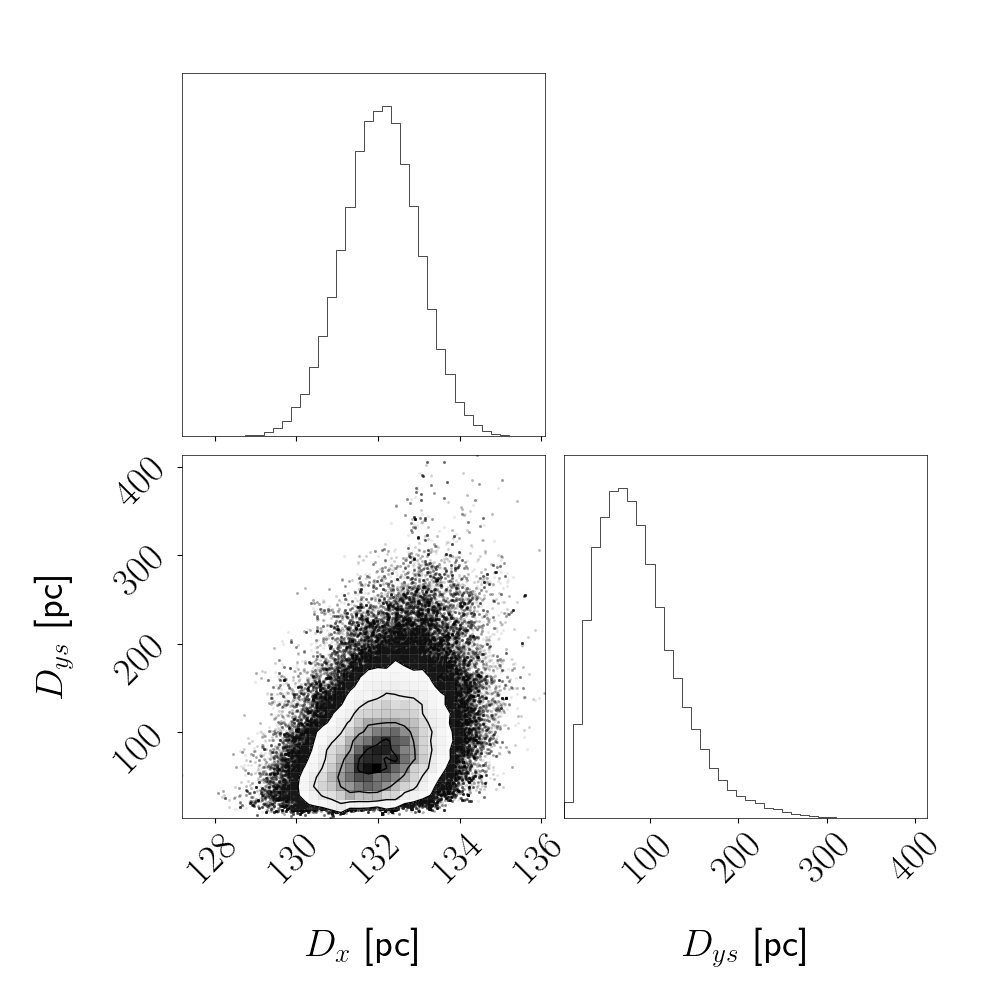}
    \caption{
    Posterior distribution from 676000 samples for the two-screen model. Left: Screen orientations at the two epochs of simultaneous observations. The red lines show the 1$\sigma$ range of the screen orientation ($\alpha_x^{SS}$) from \citet{2022MNRAS.515.6198S} and the black line marks identical angles. Right: Distance, $D_x$, of the first screen from Earth against the distance, $D_{ys}$, of the second screen from the pulsar.
    }
    \label{fig:twoscreen_results}
\end{figure}

After including the two simultaneous observations, the inferred parameters for the second screen are the distance from the pulsar of $D_{y,s}=88 \pm 47\,$pc, whose posterior does not follow a Gaussian (see \cref{fig:twoscreen_results}), the orientation of $\alpha_y=-41.7\pm3.0\,^\circ$, and the projected velocity of $V_{y,\shortparallel}=-29\pm66\,$km/s. These values do not represent tighter constraints than those previously known. The second screen is also phenomenologically absent in the cross-correlation, which is compatible with a single 1D screen which is the only model predicting nearly exact identity of the two dynamic spectra beyond the shift. However, the image modulations described by \citet{2022MNRAS.515.6198S} and corresponding horizontal features on the arcs are equally visible in these observations and clearly point to the continuous presence of the second screen.

\begin{figure}
    \centering
    \includegraphics[trim={0cm 0.5cm 11.1cm 11.3cm},clip,width=.49\linewidth]{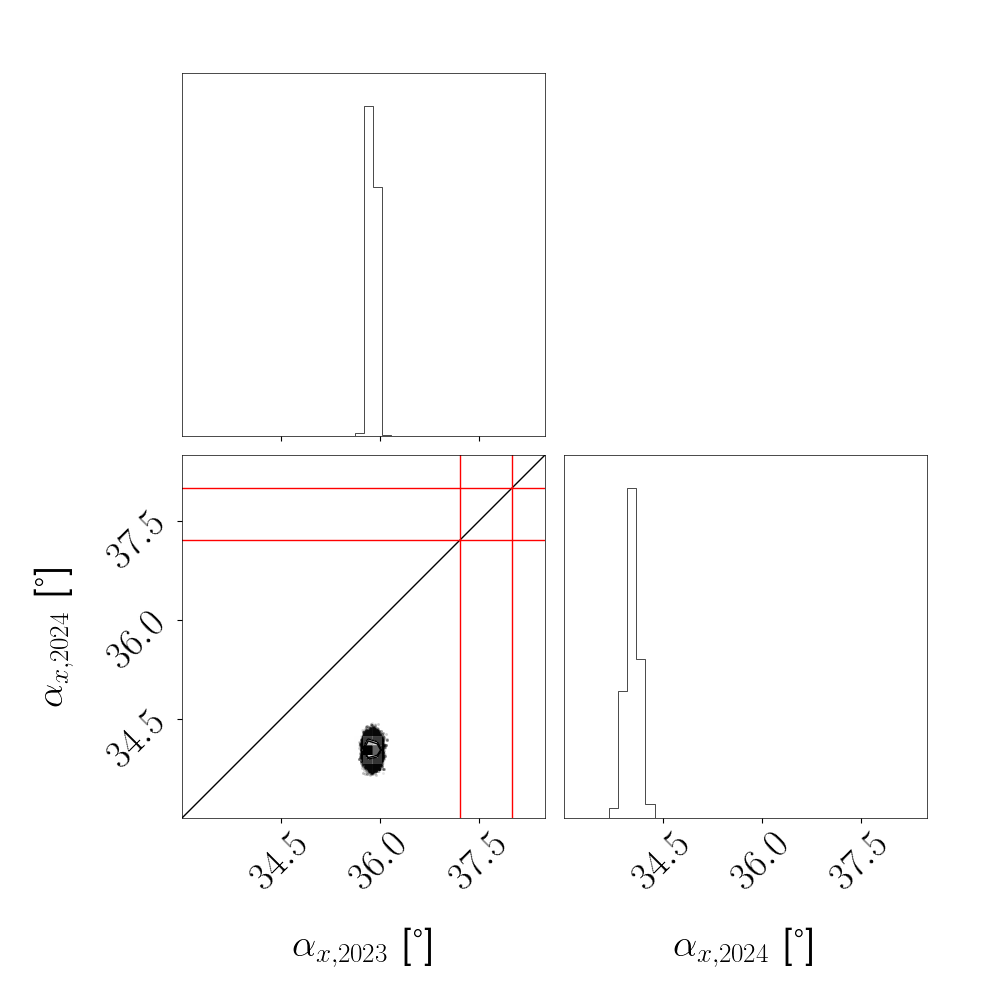}
    \includegraphics[trim={0cm 0.5cm 11.1cm 11.3cm},clip,width=.49\linewidth]{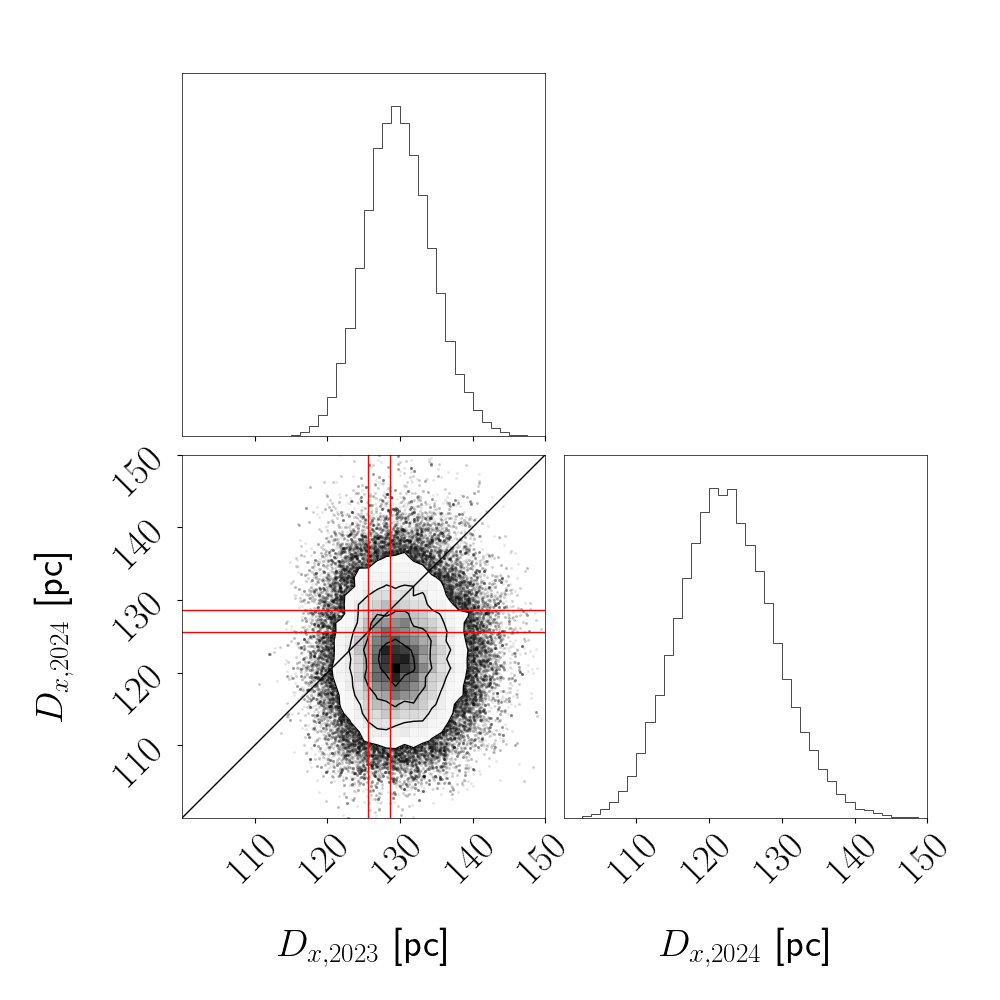}
    \caption{
    Posterior distribution from 240000 samples for the one-screen model. The red lines show the 1$\sigma$ range of the screen orientation ($\alpha_x^{SS}$) from \citet{2022MNRAS.515.6198S} and the black line marks equality of the parameters. Left: Screen orientations at the two epochs of simultaneous observations. Right: Screen distances at the same epochs.
    }
    \label{fig:onescreen_results}
\end{figure}

For comparison, we investigated a model using the new data only and setting the screen parameters orientation, distance, and velocity to be independent at each observation. Since the second screen was not constrained in this case because its shift and arc curvature were not measured, a one-screen model was employed. The same prior information on the pulsar parameters and the screen velocities was used as in the two-screen model. The results are shown in the third column of \cref{tab:mcmc_results}.

In both models, the pulsar parameters cannot be further constrained. Sampled posteriors for the screen angles and distances in the two-screen model and the one-screen model are presented in \cref{fig:twoscreen_results} and \cref{fig:onescreen_results}, respectively. The four contours go up to the $2\sigma$ region in equal steps as is the default setting by \textsc{corner} \citep{2016JOSS....1...24F}. In the one-screen model, the screen distance is compatible with being constant and with the previously inferred one, although the uncertainties are too large to exclude an evolution on the scale hinted at by the two-screen results. 

\section{Conclusions}
\label{sec:conclusion}

We performed two simultaneous observations of PSR B1508+55 at Effelsberg and with FAST. For both pairs of dynamic spectra, we formed intensity cross-secondary spectra, imaged the screen, determined the curvature parameter of scintillation arcs, and performed a cross-correlation analysis. At both epochs, the pulsar was scattered by an ISM screen into a straight line on the sky with only minor deviations. The effects of a known second screen were visible, but with a negligible impact on the aims and conclusions of this study.

We have successfully demonstrated that the scattered image of the pulsar can be imaged with \SI{0.1}{mas} resolution, which corresponds to \SI{\sim 0.01}{AU} in the ISM, without using visibilities and employing only two telescopes. This is possible because of the inherently interferometric information present in scintillation as well as the evolving projected baseline due to the rotation of the Earth and the high declination of the source. Inverted arclets are not required for this method but a distinct scintillation arc is necessary because the region along the arc needs to be identified. As a result, this method works equally for thin arcs and arcs with arclets, but will struggle with fuzzy secondary spectra and with arclet apexes too far offset from the parabolic arc. Thus, it can also be extended to cases of multiple arcs, each treated independently in a one-screen approximation. The greatly reduced logistical requirements compared to classical VLBI imaging make this a useful method for future studies of scattering screens.

The screen orientation at both epochs could be measured with a very high precision that exceeds that obtained from several years of observations of annual arc curvature variations with Effelsberg alone. This precision enabled by the long baseline and the presented cross-correlation method proves that the screen orientation is evolving between observations. The new measurements have been combined with data from \citet{2022MNRAS.515.6198S} to allow for a likelihood analysis, leading to improved constrains on the orientation, distance, and velocity of the screen.

These measurements yielded hints to an evolving distance or thickness of the closer screen on a scale of a few parsecs, while remaining in agreement with a model of two strongly scintillating, thin, 1D screens. The scattered image of the pulsar on the closer screen is about \SI{2}{AU}$\times$\SI{0.05}{AU} in terms of physical size. At a distance of around \SI{125}{pc} and a frequency of \SI{1.4}{GHz}, this indicates (in numbers) that under the stationary phase condition, the local electron column density contains gradients corresponding to a change of \SI{0.01}{pc/cm^3} in dispersion measure per astronomical unit in one direction, while at no point exceeding \SI{0.0003}{pc/cm^3} per astronomical unit in the other direction. Studies such as the one presented here can inform models of the small-scale structure in the ISM, answering the question of whether structures corresponding to 1D screens are common or, in fact, exceptional.

\begin{acknowledgements} 
Based on observations with the 100-m telescope of the MPIfR (Max-Planck-Institut für Radioastronomie) at Effelsberg. This work made use of the data from FAST (Five-hundred-meter Aperture Spherical radio Telescope).  FAST is a Chinese national mega-science facility, operated by National Astronomical Observatories, Chinese Academy of Sciences. We thank GuangXing Li and JinChen Jiang for their help with data transfer. XS thanks Jumei Yao and Weiwei Zhu for their collaboration on the earlier FAST observations of B1508+55, which laid the foundation for the simultaneous observations. XS is supported by NSFC grant No. 12373025. TS thanks Nataliya Porayko for her useful comments being the internal referee at the MPIfR.
\end{acknowledgements}

\bibliographystyle{aa}
\bibliography{references}

@ARTICLE{1968Natur.218..920S,
       author = {{Scheuer}, P.~A.~G.},
        title = "{Amplitude Variations in Pulsed Radio Sources}",
      journal = {\nat},
         year = 1968,
        month = jun,
       volume = {218},
       number = {5145},
        pages = {920-922},
          doi = {10.1038/218920a0},
       adsurl = {https://ui.adsabs.harvard.edu/abs/1968Natur.218..920S}
}

@ARTICLE{1968Natur.219..576H,
       author = {{Huguenin}, G.~R. and {Taylor}, J.~H. and {Goad}, L.~E. and {Hartai}, A. and {Orsten}, G.~S.~F. and {Rodman}, A.~K.},
        title = "{New Pulsating Radio Source}",
      journal = {\nat},
         year = 1968,
        month = aug,
       volume = {219},
       number = {5154},
        pages = {576-577},
          doi = {10.1038/219576a0},
       adsurl = {https://ui.adsabs.harvard.edu/abs/1968Natur.219..576H}
}

@ARTICLE{1973ApJ...181..875R,
       author = {{Rankin}, J.~M. and {Counselman}, III, C.~C.},
        title = "{Pulsar NP 0532: Variability of Dispersion and Scattering}",
      journal = {\apj},
         year = 1973,
        month = may,
       volume = {181},
        pages = {875-890},
          doi = {10.1086/152099},
       adsurl = {https://ui.adsabs.harvard.edu/abs/1973ApJ...181..875R}
}

@ARTICLE{1977ARA&A..15..479R,
       author = {{Rickett}, B.~J.},
        title = "{Interstellar scattering and scintillation of radio waves.}",
      journal = {\araa},
         year = 1977,
        month = jan,
       volume = {15},
        pages = {479-504},
          doi = {10.1146/annurev.aa.15.090177.002403},
       adsurl = {https://ui.adsabs.harvard.edu/abs/1977ARA&A..15..479R}
}

@ARTICLE{1985ApJ...288..221C,
       author = {{Cordes}, J.~M. and {Weisberg}, J.~M. and {Boriakoff}, V.},
        title = "{Small-scale electron density turbulence in the interstellar medium.}",
      journal = {\apj},
         year = 1985,
        month = jan,
       volume = {288},
        pages = {221-247},
          doi = {10.1086/162784},
       adsurl = {https://ui.adsabs.harvard.edu/abs/1985ApJ...288..221C}
}

@ARTICLE{1997MNRAS.287..739R,
       author = {{Rickett}, Barney J. and {Lyne}, Andrew G. and {Gupta}, Yashwant},
        title = "{Interstellar Fringes from Pulsar B0834+06}",
      journal = {\mnras},
         year = 1997,
        month = jun,
       volume = {287},
       number = {4},
        pages = {739-752},
          doi = {10.1093/mnras/287.4.739},
       adsurl = {https://ui.adsabs.harvard.edu/abs/1997MNRAS.287..739R}
}

@ARTICLE{2001ApJ...549L..97S,
       author = {{Stinebring}, D.~R. and {McLaughlin}, M.~A. and {Cordes}, J.~M. and {Becker}, K.~M. and {Goodman}, J.~E. Espinoza and {Kramer}, M.~A. and {Sheckard}, J.~L. and {Smith}, C.~T.},
        title = "{Faint Scattering Around Pulsars: Probing the Interstellar Medium on Solar System Size Scales}",
      journal = {\apjl},
         year = 2001,
        month = mar,
       volume = {549},
       number = {1},
        pages = {L97-L100},
          doi = {10.1086/319133},
archivePrefix = {arXiv},
       eprint = {astro-ph/0010363},
 primaryClass = {astro-ph},
       adsurl = {https://ui.adsabs.harvard.edu/abs/2001ApJ...549L..97S}
}

@ARTICLE{2003ApJ...599..457H,
       author = {{Hill}, Alex S. and {Stinebring}, Daniel R. and {Barnor}, Henry A. and {Berwick}, Daniel E. and {Webber}, Aaron B.},
        title = "{Pulsar Scintillation Arcs. I. Frequency Dependence}",
      journal = {\apj},
         year = 2003,
        month = dec,
       volume = {599},
       number = {1},
        pages = {457-464},
          doi = {10.1086/379191},
       adsurl = {https://ui.adsabs.harvard.edu/abs/2003ApJ...599..457H}
}

@ARTICLE{2004MNRAS.354...43W,
       author = {{Walker}, M.~A. and {Melrose}, D.~B. and {Stinebring}, D.~R. and {Zhang}, C.~M.},
        title = "{Interpretation of parabolic arcs in pulsar secondary spectra}",
      journal = {\mnras},
         year = 2004,
        month = oct,
       volume = {354},
       number = {1},
        pages = {43-54},
          doi = {10.1111/j.1365-2966.2004.08159.x},
archivePrefix = {arXiv},
       eprint = {astro-ph/0403587},
 primaryClass = {astro-ph},
       adsurl = {https://ui.adsabs.harvard.edu/abs/2004MNRAS.354...43W}
}

@ARTICLE{2005ApJ...619L.171H,
       author = {{Hill}, Alex S. and {Stinebring}, Daniel R. and {Asplund}, Curtis T. and {Berwick}, Daniel E. and {Everett}, Wendeline B. and {Hinkel}, Natalie R.},
        title = "{Deflection of Pulsar Signal Reveals Compact Structures in the Galaxy}",
      journal = {\apjl},
         year = 2005,
        month = feb,
       volume = {619},
       number = {2},
        pages = {L171-L174},
          doi = {10.1086/428347},
archivePrefix = {arXiv},
       eprint = {astro-ph/0411752},
 primaryClass = {astro-ph},
       adsurl = {https://ui.adsabs.harvard.edu/abs/2005ApJ...619L.171H}
}

@ARTICLE{2005AJ....129.1993M,
       author = {{Manchester}, R.~N. and {Hobbs}, G.~B. and {Teoh}, A. and {Hobbs}, M.},
        title = "{The Australia Telescope National Facility Pulsar Catalogue}",
      journal = {\aj},
         year = 2005,
        month = apr,
       volume = {129},
       number = {4},
        pages = {1993-2006},
          doi = {10.1086/428488},
archivePrefix = {arXiv},
       eprint = {astro-ph/0412641},
 primaryClass = {astro-ph},
       adsurl = {https://ui.adsabs.harvard.edu/abs/2005AJ....129.1993M}
}

@ARTICLE{2006ApJ...637..346C,
       author = {{Cordes}, James M. and {Rickett}, Barney J. and {Stinebring}, Daniel R. and {Coles}, William A.},
        title = "{Theory of Parabolic Arcs in Interstellar Scintillation Spectra}",
      journal = {\apj},
         year = 2006,
        month = jan,
       volume = {637},
       number = {1},
        pages = {346-365},
          doi = {10.1086/498332},
archivePrefix = {arXiv},
       eprint = {astro-ph/0407072},
 primaryClass = {astro-ph},
       adsurl = {https://ui.adsabs.harvard.edu/abs/2006ApJ...637..346C}
}

@INPROCEEDINGS{2007ASPC..365..254S,
       author = {{Stinebring}, D.},
        title = "{Pulsar Scintillation Arcs and the Ionized ISM}",
    booktitle = {SINS - Small Ionized and Neutral Structures in the Diffuse Interstellar Medium},
         year = 2007,
       editor = {{Haverkorn}, M. and {Goss}, W.~M.},
       series = {Astronomical Society of the Pacific Conference Series},
       volume = {365},
        month = jul,
        pages = {254},
       adsurl = {https://ui.adsabs.harvard.edu/abs/2007ASPC..365..254S}
}

@INPROCEEDINGS{2007ASPC..365..299W,
       author = {{Walker}, M.~A.},
        title = "{Extreme Scattering Events: Insights into the Interstellar Medium on AU-Scales}",
    booktitle = {SINS - Small Ionized and Neutral Structures in the Diffuse Interstellar Medium},
         year = 2007,
       editor = {{Haverkorn}, M. and {Goss}, W.~M.},
       series = {Astronomical Society of the Pacific Conference Series},
       volume = {365},
        month = jul,
        pages = {299},
          doi = {10.48550/arXiv.astro-ph/0610737},
archivePrefix = {arXiv},
       eprint = {astro-ph/0610737},
 primaryClass = {astro-ph},
       adsurl = {https://ui.adsabs.harvard.edu/abs/2007ASPC..365..299W}
}

@ARTICLE{2008MNRAS.388.1214W,
       author = {{Walker}, M.~A. and {Koopmans}, L.~V.~E. and {Stinebring}, D.~R. and {van Straten}, W.},
        title = "{Interstellar holography}",
      journal = {\mnras},
         year = 2008,
        month = aug,
       volume = {388},
       number = {3},
        pages = {1214-1222},
          doi = {10.1111/j.1365-2966.2008.13452.x},
archivePrefix = {arXiv},
       eprint = {0801.4183},
 primaryClass = {astro-ph},
       adsurl = {https://ui.adsabs.harvard.edu/abs/2008MNRAS.388.1214W}
}

@ARTICLE{2009ApJ...698..250C,
       author = {{Chatterjee}, S. and {Brisken}, W.~F. and {Vlemmings}, W.~H.~T. and {Goss}, W.~M. and {Lazio}, T.~J.~W. and {Cordes}, J.~M. and {Thorsett}, S.~E. and {Fomalont}, E.~B. and {Lyne}, A.~G. and {Kramer}, M.},
        title = "{Precision Astrometry with the Very Long Baseline Array: Parallaxes and Proper Motions for 14 Pulsars}",
      journal = {\apj},
         year = 2009,
        month = jun,
       volume = {698},
       number = {1},
        pages = {250-265},
          doi = {10.1088/0004-637X/698/1/250},
archivePrefix = {arXiv},
       eprint = {0901.1436},
 primaryClass = {astro-ph.SR},
       adsurl = {https://ui.adsabs.harvard.edu/abs/2009ApJ...698..250C}
}

@ARTICLE{2010ApJ...708..232B,
       author = {{Brisken}, W.~F. and {Macquart}, J. -P. and {Gao}, J.~J. and {Rickett}, B.~J. and {Coles}, W.~A. and {Deller}, A.~T. and {Tingay}, S.~J. and {West}, C.~J.},
        title = "{100 {\ensuremath{\mu}}as Resolution VLBI Imaging of Anisotropic Interstellar Scattering Toward Pulsar B0834+06}",
      journal = {\apj},
         year = 2010,
        month = jan,
       volume = {708},
       number = {1},
        pages = {232-243},
          doi = {10.1088/0004-637X/708/1/232},
archivePrefix = {arXiv},
       eprint = {0910.5654},
 primaryClass = {astro-ph.GA},
       adsurl = {https://ui.adsabs.harvard.edu/abs/2010ApJ...708..232B}
}

@ARTICLE{2011PASA...28....1V,
       author = {{van Straten}, W. and {Bailes}, M.},
        title = "{DSPSR: Digital Signal Processing Software for Pulsar Astronomy}",
      journal = {\pasa},
         year = 2011,
        month = jan,
       volume = {28},
       number = {1},
        pages = {1-14},
          doi = {10.1071/AS10021},
archivePrefix = {arXiv},
       eprint = {1008.3973},
 primaryClass = {astro-ph.IM},
       adsurl = {https://ui.adsabs.harvard.edu/abs/2011PASA...28....1V}
}

@ARTICLE{2012AR&T....9..237V,
       author = {{van Straten}, Willem and {Demorest}, Paul and {Oslowski}, Stefan},
        title = "{Pulsar Data Analysis with PSRCHIVE}",
      journal = {Astronomical Research and Technology},
         year = 2012,
        month = jul,
       volume = {9},
       number = {3},
        pages = {237-256},
          doi = {10.48550/arXiv.1205.6276},
archivePrefix = {arXiv},
       eprint = {1205.6276},
 primaryClass = {astro-ph.IM},
       adsurl = {https://ui.adsabs.harvard.edu/abs/2012AR&T....9..237V}
}

@ARTICLE{2013PASP..125..306F,
       author = {{Foreman-Mackey}, Daniel and {Hogg}, David W. and {Lang}, Dustin and {Goodman}, Jonathan},
        title = "{emcee: The MCMC Hammer}",
      journal = {\pasp},
         year = 2013,
        month = mar,
       volume = {125},
       number = {925},
        pages = {306},
          doi = {10.1086/670067},
archivePrefix = {arXiv},
       eprint = {1202.3665},
 primaryClass = {astro-ph.IM},
       adsurl = {https://ui.adsabs.harvard.edu/abs/2013PASP..125..306F}
}

@ARTICLE{2014MNRAS.442.3338P,
       author = {{Pen}, Ue-Li and {Levin}, Yuri},
        title = "{Pulsar scintillations from corrugated reconnection sheets in the interstellar medium}",
      journal = {\mnras},
         year = 2014,
        month = aug,
       volume = {442},
       number = {4},
        pages = {3338-3346},
          doi = {10.1093/mnras/stu1020},
archivePrefix = {arXiv},
       eprint = {1302.1897},
 primaryClass = {astro-ph.GA},
       adsurl = {https://ui.adsabs.harvard.edu/abs/2014MNRAS.442.3338P}
}

@ARTICLE{2014JGRA..11910544F,
       author = {{Fallows}, R.~A. and {Coles}, W.~A. and {McKay-Bukowski}, D. and {Vierinen}, J. and {Virtanen}, I.~I. and {Postila}, M. and {Ulich}, Th. and {Enell}, C. -F. and {Kero}, A. and {Iinatti}, T. and {Lehtinen}, M. and {Orisp{\"a}{\"a}}, M. and {Raita}, T. and {Roininen}, L. and {Turunen}, E. and {Brentjens}, M. and {Ebbendorf}, N. and {Gerbers}, M. and {Grit}, T. and {Gruppen}, P. and {Meulman}, H. and {Norden}, M.~J. and {de Reijer}, J. -P. and {Schoenmakers}, A. and {Stuurwold}, K.},
        title = "{Broadband meter-wavelength observations of ionospheric scintillation}",
      journal = {Journal of Geophysical Research (Space Physics)},
         year = 2014,
        month = dec,
       volume = {119},
       number = {12},
        pages = {10,544-10,560},
          doi = {10.1002/2014JA020406},
archivePrefix = {arXiv},
       eprint = {1511.00937},
 primaryClass = {astro-ph.SR},
       adsurl = {https://ui.adsabs.harvard.edu/abs/2014JGRA..11910544F}
}

@ARTICLE{2016ApJ...818...86B,
       author = {{Bhat}, N.~D.~R. and {Ord}, S.~M. and {Tremblay}, S.~E. and {McSweeney}, S.~J. and {Tingay}, S.~J.},
        title = "{Scintillation Arcs in Low-frequency Observations of the Timing-array Millisecond Pulsar PSR J0437-4715}",
      journal = {\apj},
         year = 2016,
        month = feb,
       volume = {818},
       number = {1},
          eid = {86},
        pages = {86},
          doi = {10.3847/0004-637X/818/1/86},
archivePrefix = {arXiv},
       eprint = {1512.08908},
 primaryClass = {astro-ph.SR},
       adsurl = {https://ui.adsabs.harvard.edu/abs/2016ApJ...818...86B}
}

@ARTICLE{2016JOSS....1...24F,
       author = {{Foreman-Mackey}, Daniel},
        title = "{corner.py: Scatterplot matrices in Python}",
      journal = {The Journal of Open Source Software},
         year = 2016,
        month = jun,
       volume = {1},
        pages = {24},
          doi = {10.21105/joss.00024},
       adsurl = {https://ui.adsabs.harvard.edu/abs/2016JOSS....1...24F}
}

@INPROCEEDINGS{2018evn..confE..17W,
       author = {{Wucknitz}, O.},
        title = "{Imaging pulsar echoes at low frequencies}",
    booktitle = {14th European VLBI Network Symposium \& Users Meeting (EVN 2018)},
         year = 2018,
        month = nov,
          eid = {17},
        pages = {17},
          doi = {10.22323/1.344.0017},
       adsurl = {https://ui.adsabs.harvard.edu/abs/2018evn..confE..17W}
}

@ARTICLE{2019MNRAS.485.4389R,
       author = {{Reardon}, D.~J. and {Coles}, W.~A. and {Hobbs}, G. and {Ord}, S. and {Kerr}, M. and {Bailes}, M. and {Bhat}, N.~D.~R. and {Venkatraman Krishnan}, V.},
        title = "{Modelling annual and orbital variations in the scintillation of the relativistic binary PSR J1141-6545}",
      journal = {\mnras},
         year = 2019,
        month = may,
       volume = {485},
       number = {3},
        pages = {4389-4403},
          doi = {10.1093/mnras/stz643},
archivePrefix = {arXiv},
       eprint = {1903.01990},
 primaryClass = {astro-ph.HE},
       adsurl = {https://ui.adsabs.harvard.edu/abs/2019MNRAS.485.4389R}
}

@ARTICLE{2019MNRAS.486.2809G,
       author = {{Gwinn}, Carl R.},
        title = "{Noodle model for scintillation arcs}",
      journal = {\mnras},
         year = 2019,
        month = jun,
       volume = {486},
       number = {2},
        pages = {2809-2826},
          doi = {10.1093/mnras/stz894},
archivePrefix = {arXiv},
       eprint = {1811.05996},
 primaryClass = {astro-ph.GA},
       adsurl = {https://ui.adsabs.harvard.edu/abs/2019MNRAS.486.2809G}
}

@ARTICLE{2019MNRAS.488.4952S,
       author = {{Simard}, D. and {Pen}, U. -L. and {Marthi}, V.~R. and {Brisken}, W.},
        title = "{A comparison of interferometric and single-dish methods to measure distances to pulsar scattering screens}",
      journal = {\mnras},
         year = 2019,
        month = oct,
       volume = {488},
       number = {4},
        pages = {4952-4962},
          doi = {10.1093/mnras/stz2043},
       adsurl = {https://ui.adsabs.harvard.edu/abs/2019MNRAS.488.4952S}
}

@ARTICLE{2019MNRAS.489.3692G,
       author = {{Gwinn}, Carl R. and {Sosenko}, Evan B.},
        title = "{Scintillation arc brightness and electron density for an analytical noodle model}",
      journal = {\mnras},
         year = 2019,
        month = nov,
       volume = {489},
       number = {3},
        pages = {3692-3709},
          doi = {10.1093/mnras/stz2364},
archivePrefix = {arXiv},
       eprint = {1908.00095},
 primaryClass = {astro-ph.GA},
       adsurl = {https://ui.adsabs.harvard.edu/abs/2019MNRAS.489.3692G}
}

@ARTICLE{2020ApJ...904..104R,
       author = {{Reardon}, Daniel J. and {Coles}, William A. and {Bailes}, Matthew and {Bhat}, N.~D. Ramesh and {Dai}, Shi and {Hobbs}, George B. and {Kerr}, Matthew and {Manchester}, Richard N. and {Os{\l}owski}, Stefan and {Parthasarathy}, Aditya and {Russell}, Christopher J. and {Shannon}, Ryan M. and {Spiewak}, Ren{\'e}e and {Toomey}, Lawrence and {Tuntsov}, Artem V. and {van Straten}, Willem and {Walker}, Mark A. and {Wang}, Jingbo and {Zhang}, Lei and {Zhu}, Xing-Jiang},
        title = "{Precision Orbital Dynamics from Interstellar Scintillation Arcs for PSR J0437-4715}",
      journal = {\apj},
         year = 2020,
        month = dec,
       volume = {904},
       number = {2},
          eid = {104},
        pages = {104},
          doi = {10.3847/1538-4357/abbd40},
archivePrefix = {arXiv},
       eprint = {2009.12757},
 primaryClass = {astro-ph.HE},
       adsurl = {https://ui.adsabs.harvard.edu/abs/2020ApJ...904..104R}
}

@ARTICLE{2021MNRAS.500.1114S,
       author = {{Sprenger}, Tim and {Wucknitz}, Olaf and {Main}, Robert and {Baker}, Daniel and {Brisken}, Walter},
        title = "{The {\ensuremath{\theta}}-{\ensuremath{\theta}} diagram: transforming pulsar scintillation spectra to coordinates on highly anisotropic interstellar scattering screens}",
      journal = {\mnras},
         year = 2021,
        month = jan,
       volume = {500},
       number = {1},
        pages = {1114-1124},
          doi = {10.1093/mnras/staa3353},
archivePrefix = {arXiv},
       eprint = {2008.09443},
 primaryClass = {astro-ph.IM},
       adsurl = {https://ui.adsabs.harvard.edu/abs/2021MNRAS.500.1114S}
}

@ARTICLE{2021MNRAS.506.5160M,
       author = {{Marthi}, V.~R. and {Simard}, D. and {Main}, R.~A. and {Pen}, U. -L. and {van Kerkwijk}, M.~H. and {Vanderlinde}, K. and {Gupta}, Y. and {Roberts}, C. and {Quine}, B.~M.},
        title = "{Scintillation of PSR B1508+55 - the view from a 10 000-km baseline}",
      journal = {\mnras},
         year = 2021,
        month = oct,
       volume = {506},
       number = {4},
        pages = {5160-5169},
          doi = {10.1093/mnras/stab1970},
archivePrefix = {arXiv},
       eprint = {2010.09723},
 primaryClass = {astro-ph.HE},
       adsurl = {https://ui.adsabs.harvard.edu/abs/2021MNRAS.506.5160M}
}

@ARTICLE{2022ApJ...935..167A,
       author = {{Astropy Collaboration} and {Price-Whelan}, Adrian M. and {Lim}, Pey Lian and {Earl}, Nicholas and {Starkman}, Nathaniel and {Bradley}, Larry and {Shupe}, David L. and {Patil}, Aarya A. and {Corrales}, Lia and {Brasseur}, C.~E. and {N{\"o}the}, Maximilian and {Donath}, Axel and {Tollerud}, Erik and {Morris}, Brett M. and {Ginsburg}, Adam and {Vaher}, Eero and {Weaver}, Benjamin A. and {Tocknell}, James and {Jamieson}, William and {van Kerkwijk}, Marten H. and {Robitaille}, Thomas P. and {Merry}, Bruce and {Bachetti}, Matteo and {G{\"u}nther}, H. Moritz and {Aldcroft}, Thomas L. and {Alvarado-Montes}, Jaime A. and {Archibald}, Anne M. and {B{\'o}di}, Attila and {Bapat}, Shreyas and {Barentsen}, Geert and {Baz{\'a}n}, Juanjo and {Biswas}, Manish and {Boquien}, M{\'e}d{\'e}ric and {Burke}, D.~J. and {Cara}, Daria and {Cara}, Mihai and {Conroy}, Kyle E. and {Conseil}, Simon and {Craig}, Matthew W. and {Cross}, Robert M. and {Cruz}, Kelle L. and {D'Eugenio}, Francesco and {Dencheva}, Nadia and {Devillepoix}, Hadrien A.~R. and {Dietrich}, J{\"o}rg P. and {Eigenbrot}, Arthur Davis and {Erben}, Thomas and {Ferreira}, Leonardo and {Foreman-Mackey}, Daniel and {Fox}, Ryan and {Freij}, Nabil and {Garg}, Suyog and {Geda}, Robel and {Glattly}, Lauren and {Gondhalekar}, Yash and {Gordon}, Karl D. and {Grant}, David and {Greenfield}, Perry and {Groener}, Austen M. and {Guest}, Steve and {Gurovich}, Sebastian and {Handberg}, Rasmus and {Hart}, Akeem and {Hatfield-Dodds}, Zac and {Homeier}, Derek and {Hosseinzadeh}, Griffin and {Jenness}, Tim and {Jones}, Craig K. and {Joseph}, Prajwel and {Kalmbach}, J. Bryce and {Karamehmetoglu}, Emir and {Ka{\l}uszy{\'n}ski}, Miko{\l}aj and {Kelley}, Michael S.~P. and {Kern}, Nicholas and {Kerzendorf}, Wolfgang E. and {Koch}, Eric W. and {Kulumani}, Shankar and {Lee}, Antony and {Ly}, Chun and {Ma}, Zhiyuan and {MacBride}, Conor and {Maljaars}, Jakob M. and {Muna}, Demitri and {Murphy}, N.~A. and {Norman}, Henrik and {O'Steen}, Richard and {Oman}, Kyle A. and {Pacifici}, Camilla and {Pascual}, Sergio and {Pascual-Granado}, J. and {Patil}, Rohit R. and {Perren}, Gabriel I. and {Pickering}, Timothy E. and {Rastogi}, Tanuj and {Roulston}, Benjamin R. and {Ryan}, Daniel F. and {Rykoff}, Eli S. and {Sabater}, Jose and {Sakurikar}, Parikshit and {Salgado}, Jes{\'u}s and {Sanghi}, Aniket and {Saunders}, Nicholas and {Savchenko}, Volodymyr and {Schwardt}, Ludwig and {Seifert-Eckert}, Michael and {Shih}, Albert Y. and {Jain}, Anany Shrey and {Shukla}, Gyanendra and {Sick}, Jonathan and {Simpson}, Chris and {Singanamalla}, Sudheesh and {Singer}, Leo P. and {Singhal}, Jaladh and {Sinha}, Manodeep and {Sip{\H{o}}cz}, Brigitta M. and {Spitler}, Lee R. and {Stansby}, David and {Streicher}, Ole and {{\v{S}}umak}, Jani and {Swinbank}, John D. and {Taranu}, Dan S. and {Tewary}, Nikita and {Tremblay}, Grant R. and {de Val-Borro}, Miguel and {Van Kooten}, Samuel J. and {Vasovi{\'c}}, Zlatan and {Verma}, Shresth and {de Miranda Cardoso}, Jos{\'e} Vin{\'\i}cius and {Williams}, Peter K.~G. and {Wilson}, Tom J. and {Winkel}, Benjamin and {Wood-Vasey}, W.~M. and {Xue}, Rui and {Yoachim}, Peter and {Zhang}, Chen and {Zonca}, Andrea and {Astropy Project Contributors}},
        title = "{The Astropy Project: Sustaining and Growing a Community-oriented Open-source Project and the Latest Major Release (v5.0) of the Core Package}",
      journal = {\apj},
         year = 2022,
        month = aug,
       volume = {935},
       number = {2},
          eid = {167},
        pages = {167},
          doi = {10.3847/1538-4357/ac7c74},
archivePrefix = {arXiv},
       eprint = {2206.14220},
 primaryClass = {astro-ph.IM},
       adsurl = {https://ui.adsabs.harvard.edu/abs/2022ApJ...935..167A}
}

@ARTICLE{2022ApJ...941...34S,
       author = {{Stinebring}, Dan R. and {Rickett}, Barney J. and {Minter}, Anthony H. and {Hill}, Alex S. and {Jussila}, Adam P. and {Mathis}, Lele and {McLaughlin}, Maura A. and {Ocker}, Stella Koch and {Ransom}, Scott M.},
        title = "{A Scintillation Arc Survey of 22 Pulsars with Low to Moderate Dispersion Measures}",
      journal = {\apj},
         year = 2022,
        month = dec,
       volume = {941},
       number = {1},
          eid = {34},
        pages = {34},
          doi = {10.3847/1538-4357/ac8ea8},
archivePrefix = {arXiv},
       eprint = {2207.08756},
 primaryClass = {astro-ph.HE},
       adsurl = {https://ui.adsabs.harvard.edu/abs/2022ApJ...941...34S}
}

@ARTICLE{2022MNRAS.510.4573B,
       author = {{Baker}, Daniel and {Brisken}, Walter and {van Kerkwijk}, Marten H. and {Main}, Robert and {Pen}, Ue-Li and {Sprenger}, Tim and {Wucknitz}, Olaf},
        title = "{Interstellar interferometry: precise curvature measurement from pulsar secondary spectra}",
      journal = {\mnras},
         year = 2022,
        month = mar,
       volume = {510},
       number = {3},
        pages = {4573-4581},
          doi = {10.1093/mnras/stab3599},
archivePrefix = {arXiv},
       eprint = {2101.04646},
 primaryClass = {astro-ph.IM},
       adsurl = {https://ui.adsabs.harvard.edu/abs/2022MNRAS.510.4573B}
}

@ARTICLE{2022MNRAS.511.1104M,
       author = {{Mall}, G. and {Main}, R.~A. and {Antoniadis}, J. and {Bassa}, C.~G. and {Burgay}, M. and {Chen}, S. and {Cognard}, I. and {Concu}, R. and {Corongiu}, A. and {Gaikwad}, M. and {Hu}, H. and {Janssen}, G.~H. and {Karuppusamy}, R. and {Kramer}, M. and {Lee}, K.~J. and {Liu}, K. and {McKee}, J.~W. and {Melis}, A. and {Mickaliger}, M.~B. and {Perrodin}, D. and {Pilia}, M. and {Possenti}, A. and {Reardon}, D.~J. and {Sanidas}, S.~A. and {Sprenger}, T. and {Stappers}, B.~W. and {Wang}, L. and {Wucknitz}, O. and {Zhu}, W.~W.},
        title = "{Modelling annual scintillation arc variations in PSR J1643-1224 using the Large European Array for Pulsars}",
      journal = {\mnras},
         year = 2022,
        month = mar,
       volume = {511},
       number = {1},
        pages = {1104-1114},
          doi = {10.1093/mnras/stac096},
archivePrefix = {arXiv},
       eprint = {2201.04245},
 primaryClass = {astro-ph.HE},
       adsurl = {https://ui.adsabs.harvard.edu/abs/2022MNRAS.511.1104M}
}

@ARTICLE{2022MNRAS.515.6198S,
       author = {{Sprenger}, Tim and {Main}, Robert and {Wucknitz}, Olaf and {Mall}, Geetam and {Wu}, Jason},
        title = "{Double-lens scintillometry: the variable scintillation of pulsar B1508 + 55}",
      journal = {\mnras},
         year = 2022,
        month = oct,
       volume = {515},
       number = {4},
        pages = {6198-6216},
          doi = {10.1093/mnras/stac2160},
archivePrefix = {arXiv},
       eprint = {2204.13985},
 primaryClass = {astro-ph.HE},
       adsurl = {https://ui.adsabs.harvard.edu/abs/2022MNRAS.515.6198S}
}

@ARTICLE{2023MNRAS.525..211B,
       author = {{Baker}, Daniel and {Brisken}, Walter and {van Kerkwijk}, Marten H. and {van Lieshout}, Rik and {Pen}, Ue-Li},
        title = "{High-resolution VLBI astrometry of pulsar scintillation screens with the {\ensuremath{\theta}} - {\ensuremath{\theta}} transform}",
      journal = {\mnras},
         year = 2023,
        month = oct,
       volume = {525},
       number = {1},
        pages = {211-220},
          doi = {10.1093/mnras/stad2318},
archivePrefix = {arXiv},
       eprint = {2212.01417},
 primaryClass = {astro-ph.IM},
       adsurl = {https://ui.adsabs.harvard.edu/abs/2023MNRAS.525..211B}
}

@ARTICLE{2024MNRAS.527.7568O,
       author = {{Ocker}, Stella Koch and {Cordes}, James M. and {Chatterjee}, Shami and {Stinebring}, Daniel R. and {Dolch}, Timothy and {Giannakopoulos}, Christos and {Pelgrims}, Vincent and {McKee}, James W. and {Reardon}, Daniel J.},
        title = "{Pulsar scintillation through thick and thin: bow shocks, bubbles, and the broader interstellar medium}",
      journal = {\mnras},
         year = 2024,
        month = jan,
       volume = {527},
       number = {3},
        pages = {7568-7587},
          doi = {10.1093/mnras/stad3683},
archivePrefix = {arXiv},
       eprint = {2309.13809},
 primaryClass = {astro-ph.HE},
       adsurl = {https://ui.adsabs.harvard.edu/abs/2024MNRAS.527.7568O}
}

@ARTICLE{2024MNRAS.531.3950K,
       author = {{Kramer}, Tobias and {Waltner}, Daniel and {Heller}, Eric J. and {Stinebring}, Dan R.},
        title = "{Scattering model of scintillation arcs in pulsar secondary spectra}",
      journal = {\mnras},
         year = 2024,
        month = jul,
       volume = {531},
       number = {4},
        pages = {3950-3960},
          doi = {10.1093/mnras/stae1342},
archivePrefix = {arXiv},
       eprint = {2402.19370},
 primaryClass = {astro-ph.HE},
       adsurl = {https://ui.adsabs.harvard.edu/abs/2024MNRAS.531.3950K}
}

@ARTICLE{2025NatAs...9.1053R,
       author = {{Reardon}, Daniel J. and {Main}, Robert and {Ocker}, Stella Koch and {Shannon}, Ryan M. and {Bailes}, Matthew and {Camilo}, Fernando and {Geyer}, Marisa and {Jameson}, Andrew and {Kramer}, Michael and {Parthasarathy}, Aditya and {Spiewak}, Ren{\'e}e and {van Straten}, Willem and {Venkatraman Krishnan}, Vivek},
        title = "{Bow shock and Local Bubble plasma unveiled by the scintillating millisecond pulsar J0437{\ensuremath{-}}4715}",
      journal = {Nature Astronomy},
         year = 2025,
        month = jul,
       volume = {9},
        pages = {1053-1063},
          doi = {10.1038/s41550-025-02534-6},
archivePrefix = {arXiv},
       eprint = {2410.21390},
 primaryClass = {astro-ph.HE},
       adsurl = {https://ui.adsabs.harvard.edu/abs/2025NatAs...9.1053R}
}

@ARTICLE{2025ApJ...980...80S,
       author = {{Stock}, Ashley M. and {van Kerkwijk}, Marten H.},
        title = "{Associations between Scattering Screens and Interstellar Medium Filaments}",
      journal = {\apj},
         year = 2025,
        month = feb,
       volume = {980},
       number = {1},
          eid = {80},
        pages = {80},
          doi = {10.3847/1538-4357/ada1d8},
archivePrefix = {arXiv},
       eprint = {2407.16876},
 primaryClass = {astro-ph.HE},
       adsurl = {https://ui.adsabs.harvard.edu/abs/2025ApJ...980...80S}
}

@ARTICLE{2025ApJ...992..192S,
       author = {{Stock}, Ashley M. and {Syed}, Fardin and {van Kerkwijk}, Marten H. and {Lin}, Rebecca and {Kirsten}, Franz and {Pen}, Ue-Li},
        title = "{Scintillation Properties of PSR B1133+16 Measured with Very Long Baseline Interferometry}",
      journal = {\apj},
         year = 2025,
        month = oct,
       volume = {992},
       number = {2},
          eid = {192},
        pages = {192},
          doi = {10.3847/1538-4357/ae0742},
archivePrefix = {arXiv},
       eprint = {2506.02165},
 primaryClass = {astro-ph.HE},
       adsurl = {https://ui.adsabs.harvard.edu/abs/2025ApJ...992..192S}
}

@INPROCEEDINGS{EDD,
       author = {{Barr}, Ewan D. and {Bansod}, Amid and {Behrend}, Jan and {Esser}, Niclas and {Kasemann}, Christoph and {Wieching}, Gundolf and {Winchen}, Tobias and {Wu}, Jason},
    title = "{The Effelsberg Direct Digitisation System}",
    booktitle = {2023 XXXVth General Assembly and Scientific Symposium of the International Union of Radio Science (URSI GASS)},
         year = 2023,
        month = aug,
          eid = {We-J01-PM3-4},
        pages = {We-J01-PM3-4},
          doi = {10.46620/URSIGASS.2023.1703.NURX4173},
}

\appendix

\end{document}